\documentclass{aa}

\usepackage{graphicx}
\usepackage{natbib}
\usepackage{pifont}
\usepackage{txfonts}
\usepackage{mathrsfs}
\usepackage{epsfig}

\def\st{{\rm \hspace{0.1ex}\circ\hspace{-0.9ex}-}}
\def\dG{\Delta_{\rm f}^{\st}\hspace{-0.2ex}G}

\bibpunct{(}{)}{;}{a}{}{,}      

\begin{document}
  \DeclareGraphicsExtensions{.pdf,.mps,.png,.ps,.eps,.jpg}
%
\title{Dust in brown dwarfs and extra-solar planets}
\subtitle{IV. Assessing TiO$_{2}$ and SiO nucleation for cloud formation modelling}
\author{G. Lee$^1$        
        \and Ch. Helling$^1$ 
        \and H. Giles$^1$     
        \and S. T. Bromley$^{2, 3}$ }
 \institute{$^1$ SUPA, School of Physics and Astronomy, University of St. Andrews, North Haugh, St. Andrews, KY16 9SS, United Kingdom\\ email: ch80@st-and.ac.uk\\
$^2$ Departament de Quimica Fisica and IQTCUB, Universitat de Barcelona, Marti i Franques 1, E-08028 Barcelona, Spain\\
$^3$ Institucio Catalana de Recerca i Estudis Avancats (ICREA), E-08010 Barcelona, Spain }
\date{Accepted 2012 ?,  Received 2012 ?,   in original from \today}

\abstract{Clouds form in atmospheres of brown dwarfs and planets. The
  cloud particle formation processes, seed formation and
  growth/evaporation, are very similar to the dust formation process
  studied in circumstellar shells of AGB stars and in
  Supernovae. Cloud formation modelling in substellar objects requires gravitational
  settling and element replenishment in addition to element
  depletion. All processes depend on the local conditions, and a simultaneous treatment is required.}
{We apply new material data in order to assess our cloud formation
  model results regarding the treatment of the formation of condensation
  seeds. We re-address the question of the primary nucleation species
  in view of new (TiO$_{2}$)$_{N}$-cluster data and new SiO vapour
  pressure data. }
{We apply the density functional theory (B3LYP, 6-311G(d)) using the computational chemistry package {{\sc Gaussian 09}}  to derive
  updated thermodynamical data for (TiO$_{2}$)$_{N}$-clusters as input
  for our TiO$_2$ seed formation model. We test different nucleation
  treatments and their effect on the overall cloud structure by
  solving a system of dust moment equations and element conservation
  for a pre-scribed {{\sc Drift-Phoenix }} atmosphere structure. }
{Updated Gibbs free energies for the (TiO$_{2}$)$_{N}$-clusters are
  presented, and a slightly temperature dependent surface tension for
  T=500$\,\ldots\,$2000K with an average value of $\sigma_{\infty}$ =
  480.6 erg cm$^{-2}$. The TiO$_2$-seed formation rate changes only
  slightly with the updated cluster data. A considerably larger effect
  on the rate of seed formation, and hence on grain size and dust
  number density, results from a switch to SiO-nucleation. The
  question about the most efficient nucleation species can only be
  answered if all dust/cloud formation processes and their feedback
  are taken into account. Despite the higher abundance of SiO over
  TiO$_2$ in the gas phase, TiO$_2$ remains considerably more
  efficient in forming condensation seeds by homogeneous
  nucleation. The paper discussed the effect on the cloud structure in
  more detail. }{} 

\maketitle


\section{Introduction}

Brown dwarfs have long been known to form dust in atmospheres and
recent detections demonstrate their observational comparability to
giant exopanets like for 2M0355 and 2M1207b (see Faherty et al. 2013).
Transit spectroscopy observations of exoplanets suggest the presence
of haze layers in HD~189733b (Lecavelier Des Etangs et al. 2008, Sing
et al. 2011, Pont et al. 2013), GJ~1214b (Miller-Ricci Kempton et
al. 2012), WASP-12b (Copperwheat et al. 2013, Sing et al. 2013),
CoRot-1b (Schlawin et al. 2014).  The formation of cloud particles
impacts the observed spectrum of all of these objects by depleting the
local gas phase and by providing an additional opacity source. The
interpretation of such observations requires understanding and
modelling of the cloud formation processes. We will demonstrate that
the processes involved in cloud formation can not be treated
independently {\it a priori}, instead their interacting feedback needs
to be considered. This includes formation of new particles
(nucleation), the growth and evaporation of existing particle, their
gravitational settling (or other large scale relative motions),
convective mixing, and element depletion.

Nucleation rates of various chemical species are important for the formation of cloud layers, but also for modelling the element enrichment  by winds  of AGB stars and Supernovae.  In oxygen rich
atmospheres TiO$_{2}$ molecules have been identified as a key players
in seed formation due to its chemically reactive sites. In
addition, the stability of TiO$_{2}$[s] has been proven
experimentally (Demyk et al 2004) which further supports it as a
likely candidate for nucleation seeds.  

Previous work on (TiO$_{2}$)$_{N}$ clusters as precursors for
condensation seeds, that form through a step-wise increase of cluster
size, in astrophysics comes mostly from Jeong et al (2000 \& 2003) who
investigated (TiO$_{2}$)$_{N}$ nucleation in pulsating AGB stars.
They computed the most probable cluster geometries for
N~$=1\,\ldots\,6$ and recommend a surface tension value of
$\sigma_{\infty}$ = 618 erg cm$^{-2}$.  Since then, more stable,
(TiO$_{2}$)$_{N}$ cluster geometries up to $N = 10$ have been
published (e.g Calatayud et al 2008, Syzgantseva et al 2010).  Efforts
to link these small scale nano regime properties to the large scale
micron sized bulk properties and vice versa has been undertaken by
Bromley et al (2009) which noted problems in acquiring stable
TiO$_{2}$ nano-cluster geometries.  In the present paper, we use these
cluster geometries from the chemistry literature and compute Gibbs
formation energies for these clusters using the {\sc GAUSSIAN} package
(Frisch et al. 2009; Sects.~\ref{sec:method},~\ref{ThETiO2}), and then
update the surface tension value. After demonstrating the relative
abundances of the individual clusters (Sect.~\ref{sec:molabund}), we
assess the results for seed formation rates resulting from the
classical nucleation theory and from directly applying the cluster
data (Sec.~\ref{sec:J*many}).  We note that the need for calculating a
seed formation rate arrises from our kinetic treatment of cloud
particle formation. Other authors chose to treat the cloud particles
as in phase-equilibrium. For a comparison of these approaches, please
refer to Helling et al. (2008). Section~\ref{ssec:SiOnuc} compares the
TiO$_2$ seed formation rates with SiO nucleation for which updated
vapour pressure data is available. Section~\ref{sec:cloud}
demonstrates the influence of the nucleation data on the details of
the cloud structure. We show that the question regarding the most
suitable nucleation species can not be answered without taking into
account the surface growth (or evaporation) processes as they reduce
(or enrich) the gas reservoir from which the seed particles form.

\noindent 
\section{Modelling seed formation as first step of astrophysical dust and cloud formation}

Cloud formation in brown dwarfs and planets as well as dust formation
in AGB stars and Supernovae require knowledge of how the individual
(cloud) particles/grains form. The very first process is the formation
of condensation seeds, unless seeds are injected into a condensible gas like on
Earth or into the ISM through Supernovae and AGB star winds. Only the
presence of condensation seeds allows the growth to massive
($\mu$m-sized) particles (dust grains or cloud particles). Recent
developments in computational chemistry and progress in laboratory
astrophysics allow for the  assessment of the  seed formation modelling as in Helling \&
Woitke who apply the modified classical nucleation theory to model the
homogeneous nucleation of TiO$_2$ condensation seeds. Based on updated
dust data, we further assess the impact of the nucleation
treatment on results of our cloud formation model in Sect.~\ref{sec:cloud}.

\subsection{Nucleation Theory}
 We only summeris essential steps and
 definitions needed for this paper. We refer the reader to Helling \&
 Fomins (2013) and Gail \& Sedlmayr (2014) for further background
 reading.
\subsubsection{Classical Nucleation Theory}\label{ssec:clnucT}
The stationary  rate for a homogeneous, homomolecular nucleation process is given by
\begin{equation}
 J_{*}^{c}(t) = \frac{\stackrel{\circ}{f}(N_{*})}{\tau_{gr}(r_{i}, N_{*}, t)} Z(N_{*}) S(T) \cdot \exp{\{(N_{*} - 1) \ln{S(T)}\}}
\label{eq:J*}
\end{equation}
with N$_{*}$  the critical cluster size (see Eq.~\ref{eq:N*}).
The equilibrium cluster size distribution, $\stackrel{\circ}{f}(N)$ [cm$^{-3}$],  can be considered as a Boltzmann-like distribution in local thermal equilibrium,
\begin{equation}
 \stackrel{\circ}{f}(N) = \stackrel{\circ}{f}(1) \exp{\left( - \frac{\Delta G(N)}{RT}\right)},
\label{eq:f(N)}
\end{equation}
where $\stackrel{\circ}{f}$(1) [cm$^{-3}$] is the equilibrium number
density of the monomer (smallest cluster unit like TiO$_2$ or SiO) and
$\Delta G(N)$ [kJ mol$^{-1}$] is the Gibbs free energy change due to the
formation of cluster of size $N$ from the saturated vapor at
temperature $T$.  The rate of growth for each individual cluster of
size $N$ is 
\begin{equation}
\tau^{-1}_{\rm gr}(r_{\rm i}, N, t) = A(N) \alpha(r_{\rm i}, N) v_{\rm rel}(n_{\rm f}(r_{\rm i}),N) n_{\rm f}(r_{\rm i},t),
\label{eq:taugr}
\end{equation}
where $A(N)= 4\pi a_{0}^{2} N^{2/3}$ [cm$^2$] is the reaction surface area of a $N$-sized
cluster, $N$ is the number of monomers in a cluster, a$_{0}$ the  hypothetical
monomer radius,  $\alpha$ is the efficiency of the reaction (assumed to be 1);
v$_{\rm rel}$ [cm$^2$s$^{-1}$] is the relative velocity between a
monomer and the cluster, and n$_{\rm f}$ [cm$^{-3}$] is the particle density of the
molecule for the growth (forward) reaction ($\equiv\stackrel{\circ}{f}$(1)). 
The relative velocity is approximated by the thermal velocity (see Eq. 15 in Helling et al. 2001)
\begin{equation}
v_{\rm rel} = \sqrt{\frac{kT}{2\pi \bar{\mu}}} \approx \sqrt{\frac{kT}{2\pi m_{\rm x}}}
\end{equation}
with $\bar{\mu}= 1/(1/m_x - 1/m_V)$, where m$_x$ is the mass of the
momomer molecule (e.g. TiO$_{2}$) and m$_V$ the mass of a grain with
volume $V$. For macroscopic grains, m$_V$ $>>$ m$_x$, hence
$\bar{\mu}\approx m_{\rm x}$.  Equation~\ref{eq:J*} also contains the
supersaturation ratio S$_{N}$ of a cluster with $N$ monomers $S_{N} =
{\stackrel{\circ}{p}(N)}/{\stackrel{\circ}{p}_{sat}}$ with the
saturation vapour pressure
\begin{equation}
 \frac{\stackrel{\circ}{p_{sat}}}{p^{\st}} = \exp\left({\frac{\dG_{1}(s) - \dG(1)}{R T_{d}}}\right).
\label{eq:pvap}
\end{equation}
$\Delta G$(N) can be expressed by a relationship to the standard molar
Gibbs free energies in reference state ''${\rm \hspace{0.1ex}\circ\hspace{-1.25ex}-}$'' (measured at a
standard gas pressure and gas temperature) of formation for cluster
size $N$
\begin{equation}
 \Delta G(N) = \dG(N) + RT \ln\left(\frac{\stackrel{\circ}{p_{sat}}(T)}{p^{\st}}\right) - N\dG_{1}(s).
\label{eq:DeltaG}
\end{equation}
Combining  Eqs.~\ref{eq:pvap} and~\ref{eq:DeltaG} results in
\begin{equation}
 \Delta G(N) = \dG(N) - \dG(1) - (N - 1)\dG_{1}(s)
\end{equation}
where the right hand side contains standard state values only
($\dG$(N) - standard Gibbs free energy of formation of cluster size
$N$, $\dG$(1) - standard Gibbs free energy of the monomer,
$\dG_{1}$(s) - standard Gibbs free energy of formation of the solid
phase) which can be found by experiment or computational chemistry.  We
use the JANAF thermochemical table (Chase et al. 1985) where the standard
states are given at $T^{\st}_{\rm gas}$ = 298.15K and $p^{\st}_{\rm gas} = 1$
bar.

The classical nucleation theory assumes that the detailed knowledge
about $\Delta G$(N) can be encapsulated by the value of the surface
tension, $\sigma_{\infty}$, of the macroscopic solid such that
\begin{equation}
 \frac{\Delta G(N)}{RT} = -N \ln(S) + \theta_{\infty} N^{2/3} \quad \mbox{with} \quad  \theta_{\infty} = \frac{4\pi a_{0}^{2} \sigma_{\infty}}{k_{b}T}.
\label{eq:dGclass}
\end{equation}
The dependence of the surface energy on cluster size is therefore neglected.
The Zeldovich factor (contribution from Brownian motion to nucleation rate) in Eq.~\ref{eq:J*} is 
\begin{equation}
 Z(N_{*}) = \left( \frac{\theta_{\infty}}{9 \pi(N_{*} - 1)^{4/3}}\frac{(1+2(\frac{N_{f}}{N_{*} - 1})^{1/3})}{(1 + (\frac{N_{\rm f}}{N_{*} - 1})^{1/3})^{3}} \right)^{1/2}.
\end{equation}
The nucleation rate can now be expressed as 
\begin{equation}
 J_{*}^{c}(t) = \frac{\stackrel{\circ}{f}(1,t)}{\tau_{gr}(1,N_{*},t)} Z(N_{*}) \exp\left((N_{*} - 1) \ln S(T) - \frac{\Delta G(N_{*})}{RT}\right).
\label{eq:J*final}
\end{equation}

\subsubsection{Modified classical nucleation theory}

Modified nucleation theory was proposed by Draine et al. (1977) and
Gail et al. (1984) by taking into account the curvature on the surface
energy for small clusters (Gail et al. 1984). Equation~\ref{eq:dGclass} changes to
\begin{equation}
 \frac{\Delta G(N)}{RT} = \theta_{\infty} \frac{N-1}{ (N-1)^{1/3} + N_{f}^{1/3}} 
 \label{eq:DGtheta}
\end{equation}
where N$_{\rm f}$ is a fitting factor representing the particle size
at which the surface energy is reduced to half of the bulk value.
This fitting factor allows to calculate a critical cluster N$_{*}$ as
\begin{equation}
 N_{*} - 1 = \frac{N_{*,\infty}}{8}\left( 1 + \sqrt{1 + 2\left(\frac{N_{f}}{N_{*,\infty}}\right)^{1/3}} - 1\left(\frac{N_{f}}{N_{*,\infty}}\right)^{1/3}\right)^{3}
 \label{eq:N*}
\end{equation}
with
\begin{equation}
 N_{*,\infty} = \left(\frac{\frac{2}{3}\theta_{\infty}}{\ln S(T)}\right)^{3}.
\end{equation}

\subsubsection{Non-classical Nucleation Theory}

If cluster data are available, $J_{*}$ can  be calculated using
cluster number densities, growth rates and evaporation rates of each
cluster size as $J_*$ is a flux through cluster space,
\begin{equation}
 J_{*}(N) = \frac{\stackrel{\circ}{f}(N-1)}{\tau_{gr}(N-1)} - \frac{\stackrel{\circ}{f}(N)}{\tau_{ev}(N)}.
\end{equation}
Applying the Becker-D{\"o}ring method (see Gail \& Sedlmayr 2014),
$f(2)$ can be eliminated from the $N=2$ equation using the $N=3$ equation,
then again for $f(3)$ and so on, resulting in the summation
\begin{equation}
 J_{*}^{-1}(t) = \sum_{N=1}^{N_{max}} \left( \frac{\tau_{gr}(r_{i}, N, t)}{\stackrel{\circ}{f}(N)} \right)
\label{eq:J*nclass}
\end{equation}
with Eq.~\ref{eq:taugr} for $\tau_{gr}$. The number density of  a cluster of size $N$ is 
\begin{equation}
\stackrel{\circ}{f}(N) = \frac{\stackrel{\circ}{p}(N)}{kT}.
\end{equation}
The partial pressures can be calculated from the law of mass action applied to a N-cluster,
\begin{equation}
\stackrel{\circ}{p}(N) = p^{\st}\left(\frac{\stackrel{\circ}{p}(1)}{p^{\st}}\right)^{N}\exp\left({-\frac{\dG(N) - N\dG(1)}{RT}}\right),
\label{eq:p(n)}
\end{equation}
with $\stackrel{\circ}{p}$(1) [dyn cm$^{-2}$] the partial pressure of
the monomer number density.  $\stackrel{\circ}{p}$(1) will be calculated is LTE
allowing the application of our equilibrium chemistry routine (Sect.~\ref{sec:equilibrium-chem}).


\subsection{Approach}\label{sec:method}
\begin{figure}
 \includegraphics[scale=0.35]{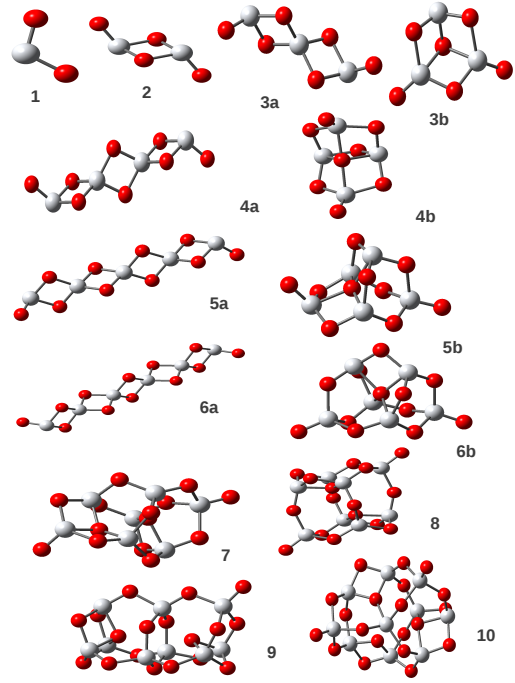}
 \caption{Geometry of the (TiO$_{2}$)$_N$ structures
   calculated. Molecules with label "a" are the molecules calculated
   by Jeong et al(2000) and those labeled "b" or unlabeled are the
   current most stable cluster geometries (Calatayud et al 2008,
   Syzgantseva et al 2010). Silver/grey balls represent Ti atoms while red represent O atoms.}
\label{fig:TiO2clusters}
\end{figure}
\subsubsection{Computational Aspects}

All cluster calculations were performed using the B3LYP (Lee et al.1988)
density functional theory with basis set 6-311G(d) as part of the \textsc{Gaussian} (Frisch et al. 2009) computational chemistry package. This level of theory was used for its mix of accuracy and computational speed as
well as keeping in line the previous investigations on the same molecules (Jeong et al. 2000, Calatayud et al. 2008, Syzgantseva et al.2011). The B3LYP functional is a popular and well regarded
functional for metal oxides and other inorganic compounds. The
reference state of the clusters was at temperature $T^{\st} =$
298.15K and pressure $p^{\st} =$1 bar. These were chosen so
that the JANAF thermochemical tables for the elemental thermochemical
values could be used for calculating the molecules
Gibbs free energies. Gaussian calculates the partition function
of a molecule using thermodynamical laws 
with contributions from the rotational, translational, vibrational and electronic motions of the molecule. Therefore, it can
generate enthalpies ($\Delta\,H$) and Gibbs energies ($\Delta\,G$) for any molecule to a
good degree of accuracy dependent only on the functional and
basis set used. These enthalpies and Gibbs free energies can then be
used to find the formation energies of the molecules with basic
thermodynamics (Sect.~\ref{ThETiO2}). Previous studies on these TiO$_{2}$ cluster geometeries  (Calatayud et al. 2008, Syzgantseva et al.2011) have focused on the reactivity and electronic structure of the clusters and not specifically the themodynamics of the formation of the clusters themselves.

\subsubsection{Cluster Geometries}
The main difference between past investigations and our research is
the calculate cluster geometries.  Figure~\ref{fig:TiO2clusters}
summarises both, the original geometries from Jeong et al. (2000)
labeled "a", and the new results labeled "b" or unlabeled.  These
geometries can mostly be found in the chemistry literature (Calatayud
et al 2008, Syzgantseva et al 2010, Richard et al 2010) except for a
new stable N=7.

The linear, polymer-like chains, investigated in Jeong et al. (2000)
(TiO$_{2}$)$_N$, are less stable than their more compressed
counterparts published by Calatayud et al 2008 and Syzgantseva et al
2010.  This is shown by the higher binding energies for the
compressed structures (Appendix A). These binding energies have a
direct impact on the Gibbs formation energies of the clusters there
will be significant differences between the two geometries.
Furthermore, it is assumed that over time the molecules will configure
to their lowest energy state geometry and so other, less stable,
configurations are therefore not considered further.


\subsection{Results for thermodynamic quantities for TiO$_{2}$ clusters}\label{ThETiO2}
Applying the results of the computation to thermodynamical identities allows the calculation of the Gibbs free energies of the clusters.
The Gibbs energy of formation can be calculated from 
\begin{equation}
\Delta_{\rm f} G^{\circ}(M,T) = \Delta_{\rm f} H^{\circ} (M,T) - T\left(S^{\circ}(M, T) -\!\!\!\!\!\sum_{\rm elements} xS^{\circ}(X, T)\right),
\end{equation}
where $M$ is the molecular/cluster values and $X$ the constituent atoms.
In order to find the enthalpy of formation of a cluster at temperature T the enthalpy of formation at 0K must first be calculated. This is given by:
\begin{equation}
 \Delta_{f}H^{\circ}(M,0K) = \sum_{atoms} x\Delta_{f}H^{\circ}(X,0K) - \sum D_{0}(M).
\label{eq:H}
\end{equation}
x is the total number of element X in the molecule and $D_{0}(M)$ the reduced atomization energy of the molecule.
The \(\Delta_{f}H^{\circ}(X,0K)\) of Ti and O can be found in the JANAF thermochemical tables.
The reduced atomization energy in Eq.~\ref{eq:H} is defined as
\begin{equation}
 \sum D_{0}(M) = \left(\sum_{elements} x E_{0}(X) - E_{0}(M) - E_{zpe}(M)\right),
\end{equation}
where $E_{0}(X)$ and $E_{0}(M)$  are the internal energy of the elements and the
molecule respectively and $E_{zpe}(M)$ the zero-point energy of the
molecule.  All the total energy terms ( \(E_{0}(X)\), $E_{0}(M)$ and
$E_{zpe}(M)$) can be calculated from the {\sc Gaussian 09} output.
The crucial quantity of the elemental atomization
energies\((E_{0}(X))\) of both Ti and O was computed using the same
level of theory (B3LYP 6-311G(d)) as the clusters.

When the enthalpy of formation at 0K is calculated for each cluster,
we can find the Enthalpy of formation at a reference temperature ($T^{\st}$ = 298.15K) as
\begin{eqnarray}
 \Delta_{f}H^{\circ}(M,298K) &=& \Delta_{f}H^{\circ}(M,0K) +  (H^{\circ}_{M}(298K) - H^{\circ}_{M}(0K)) \nonumber \\
 &&- \sum_{elements}x(H^{\circ}_{X}(298K) - H^{\circ}_{X}(0K)).
\end{eqnarray}
The enthalpy of formation at arbitrary temperature T can then be found by a similar calculation:
\begin{eqnarray}
 \Delta_{f}H^{\circ}(M,T) = \Delta_{f}H^{\circ}(M,298K) + (H^{\circ}_{M}(T) - H^{\circ}_{M}(298K)) \nonumber \\
 - \sum_{elements}x(H^{\circ}_{X}(T) - H^{\circ}_{X}(298K)).
\end{eqnarray}
The entropy of the clusters are calculated from the relation \(S =
(H - G)/T\) where H and G are the Enthalpy and Gibbs energies
respectively.  The entropy of the constituent elements at various
temperatures are  from the JANAF thermochemical tables.

Calculated thermochemical tables and Gibbs free energies for the
(TiO$_{2}$)$_N$ clusters are provided in Appendix~\ref{appB}.

\begin{figure}
 \includegraphics[scale=0.1]{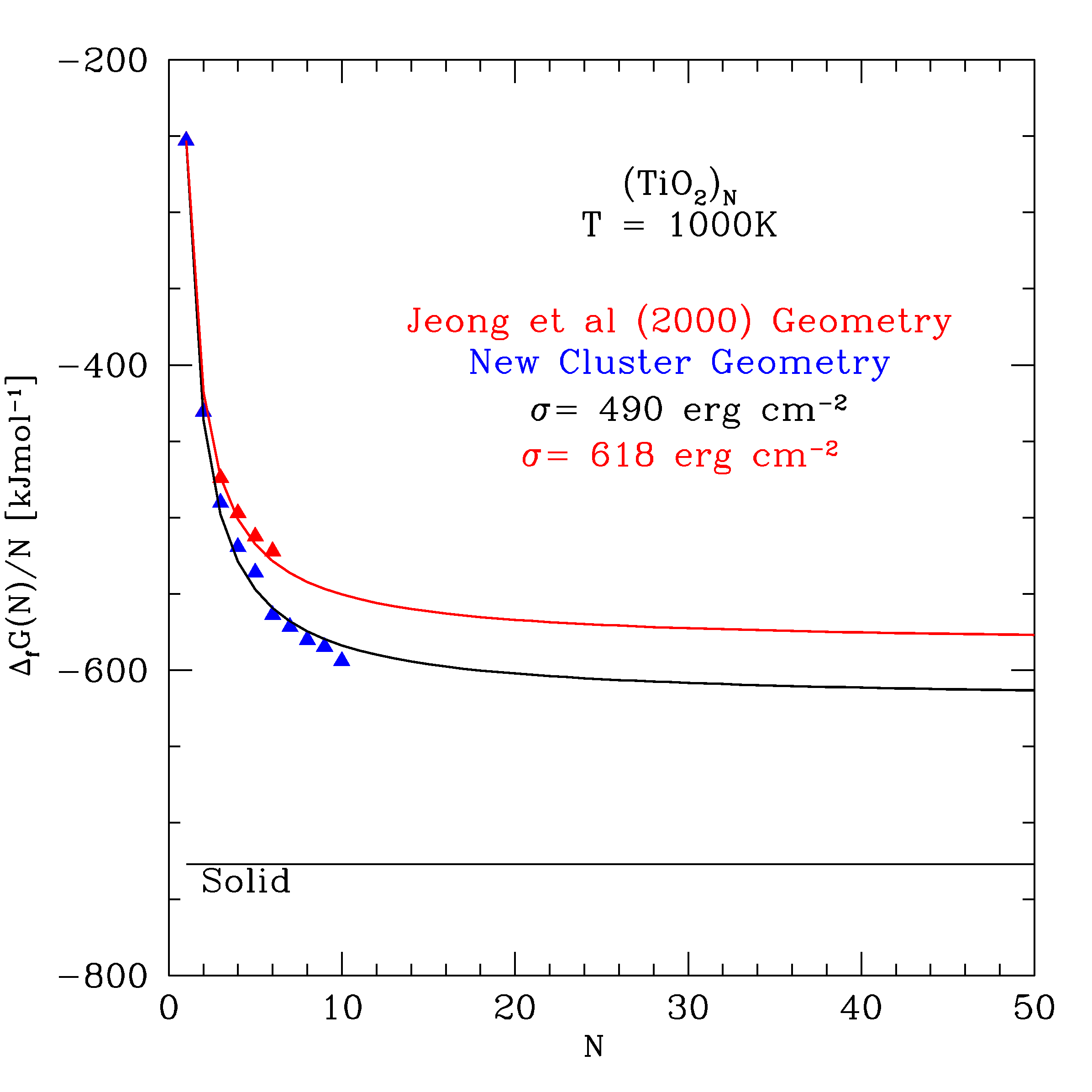}
 \caption{TiO$_{2}$ $\Delta_{f}$ G(N)/ N with respect to N for T =
   1000K. The blue triangles represent the new geometry isomers while
   red represent the molecules found in Jeong et al. (2003).  The red
   line is the modified expression with $\sigma_{\infty}$ = 618 erg
   cm$^{-2}$ and the black with $\sigma_{\infty}$ = 490 erg cm$^{-2}$}
\label{fig:G(N)}
\end{figure}

\begin{figure}
 \includegraphics[scale=0.1]{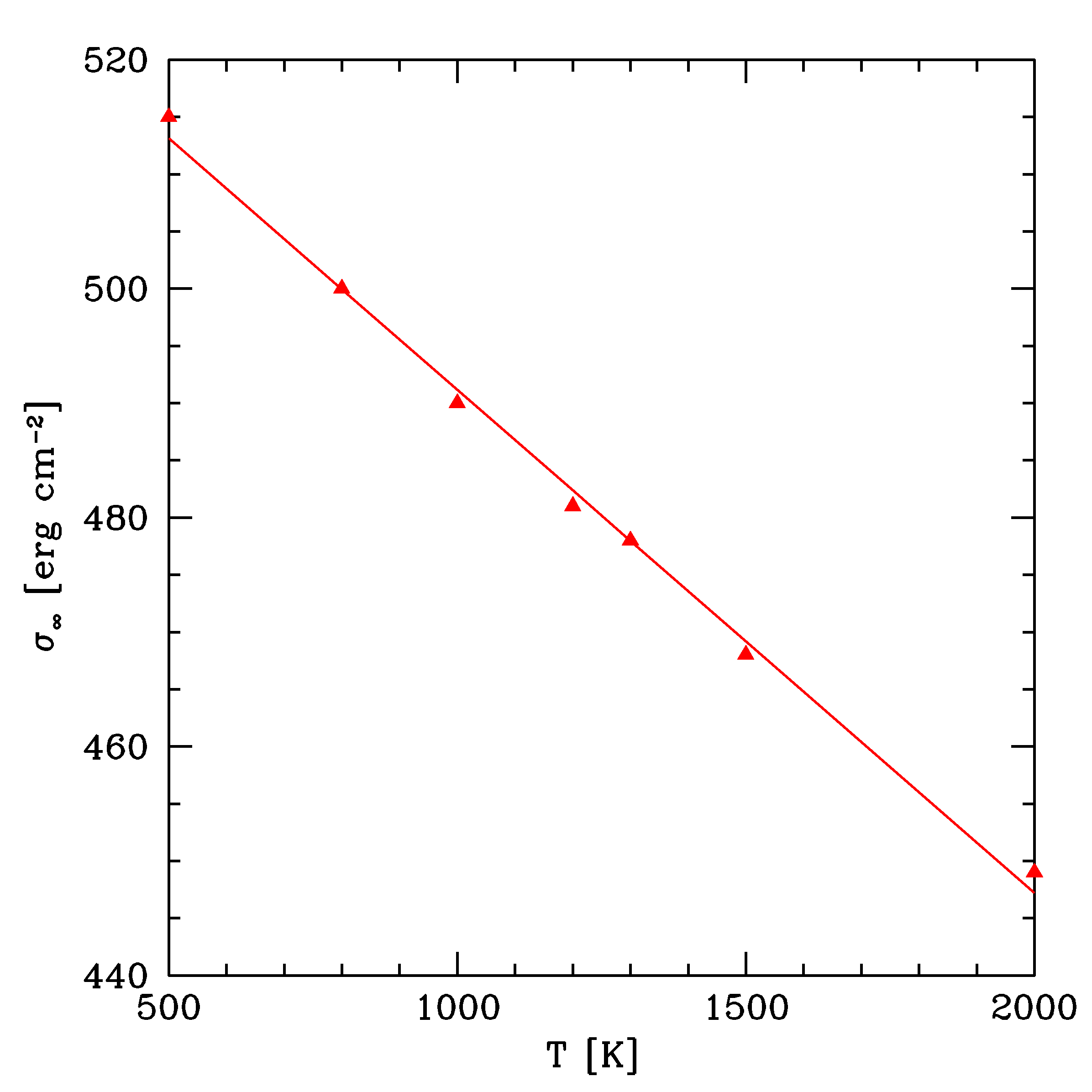}
 \caption{Temperature dependence of best fit $\sigma_{\infty}$ for the
   range $T_{\rm gas} =$ 500 - 2000 K. Triangles denote best fit values to the
   modified nucleation expression.}
\label{fig:TiO2surften}
\end{figure}

\subsubsection{Surface Tension of TiO$_{2}$}\label{ssec:SurfTen}

The surface tension is a measure of surface energy density of
the bulk property of a solid. We approximate the bulk surface
tension, $\sigma_{\infty}$,  by fitting the small clusters to the modified nucleation theory using the calculated Gibbs free energies. Combining Eqs.~\ref{eq:DeltaG} and ~\ref{eq:DGtheta},   the Gibbs formation energy of cluster size $N$ is

\begin{equation}
 \frac{\Delta_{f} G(N)}{N} = \theta_{\infty} RT \frac{N - 1}{(N-1)^{1/3} + N_{f}^{1/3}} + \Delta^{\st}_{f}G(1) + (N-1)\Delta^{\st}_{f}G_1(s).
\end{equation}
By plotting $\Delta G_{f}(N)/N$ against $N$ for the clusters, a best fit $\sigma_{\infty}$ can be found for different temperatures. Figure~\ref{fig:G(N)} shows this fitting process for T = 1000 K. The  original surface tension value from Jeong et al. (2000),  $\sigma_{\infty}$ = 618 erg cm$^{-2}$,  is also shown for comparison. The values for $\Delta^{\st}_{f}G_1(s)$  are from the JANAF table, and  N$_{f}$ = 0 is used for all calculations. The new cluster geometries have a lower Gibbs energy of
formation than the old clusters as a consequence of their increased stability. By fitting $\sigma_{\infty}$ we show that there is a slight temperature
dependentance (Fig.~\ref{fig:TiO2surften}) on the best fit value. In the range T$_{\rm gas}$ =
500 - 2000K the surface tension can be approximated by the linear relationship
\begin{equation}
 \sigma_{\infty}(T_{\rm gas}) = 535.124 - 0.04396 T_{\rm gas}.
\end{equation}
The mean value over this temperature range yields an approximate
surface tension of $\sigma_{\infty}$ = 480.6 erg cm$^{-2}$.

\section{The abundances of molecules and clusters in the gas phase}\label{sec:molabund}

The seed formation rates depend on the gas-phase composition and the
abundance of the monomer gas-species in comparison.  We therefore
summarise the abundances of the Ti-binding gas-species and we include
Si-binding molecules for later considerations of SiO-nucleation based
on updated SiO vapour pressure data.  We apply our thermodynamic
cluster data to explore the abundance of the TiO$_2$ clusters shown in
Fig.~\ref{fig:TiO2clusters} and their relative changes.

\subsection{Approach}\label{ss:DF}

We utilize one example model atmosphere structure ($T_{\rm
  eff}=1600$K, log(g)=3.0, solar metallicity) from the {\sc
  Drift-Phoenix} atmosphere grid that is representative for the
atmosphere of a giant gas planet and for brown dwarfs. This
combination of global parameters includes also the atmosphere of the group
of recently discovered low-gravity brown dwarfs (Faherty et al. 2013).
We use the ($T_{\rm gas}$, $p_{\rm gas}$) model structure as input for
our (external) chemical equilibrium program to calculate the chemical
gas composition in some more detail that necessary for the {\sc
  Drift-Phoenix} models.

\subsubsection{Drift-Phoenix model atmosphere}

{\sc Drift-Phoenix} (Dehn 2007, Helling et al. 2008b, Witte et al. 2009) model
atmosphere simulations solve the classical 1D model atmosphere
problem coupled to a kinetic phase-non-equilibrium cloud formation
model. Each of the model atmospheres is determined by the effective
temperature (T$_{\rm eff}$ [K]), the surface gravity (log(g) (with g
in cm/s$^2$), and element abundances. The cloud's opacity is
calculated applying Mie and effective medium theory.

Beside details of the dust clouds like height-dependent grain sizes
and the height-dependent composition of the mixed-material cloud
particles, the atmosphere model provides us with atmospheric
properties such as the local convective velocity,  the
temperature-pressure (T$_{\rm gas}$ [K], p$_{\rm gas}$ [dyn/cm$^2$])
structure and the dust-depleted element abundances. The local temperature is the
result of the radiative transfer solution, the local gas pressure of
the hydrostatic equilibrium, and the element abundances are the result
of the element conservation equations that include the change of
elements by dust formation and evaporation.

\begin{figure}
 \includegraphics[scale=0.35]{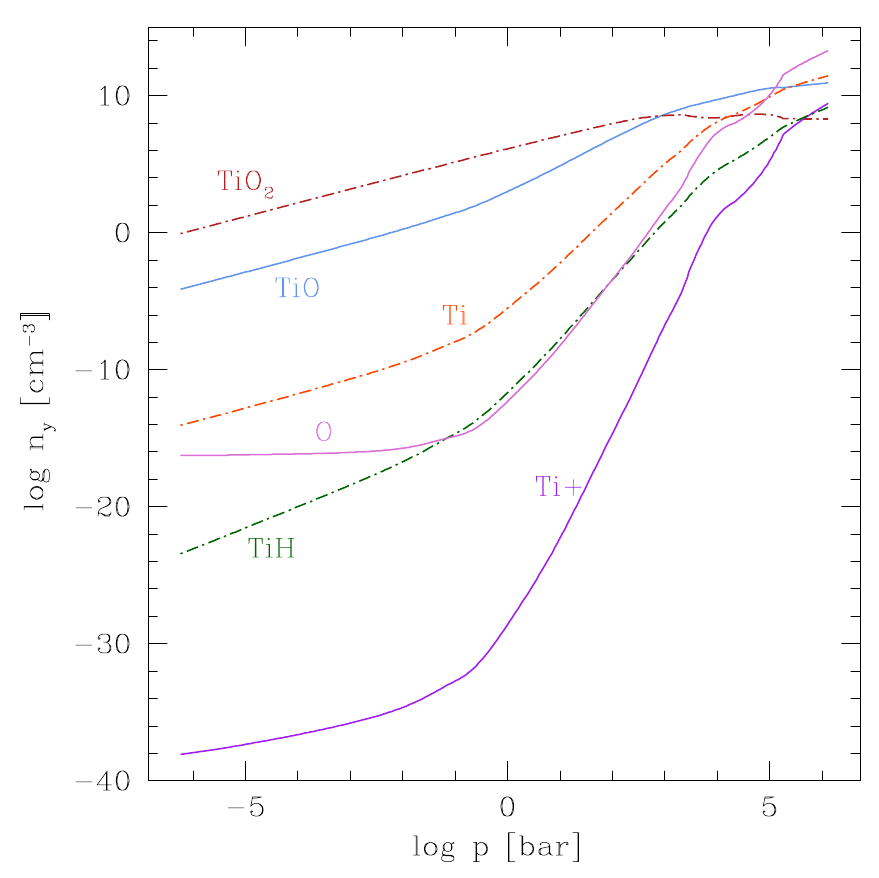}\\
 \includegraphics[scale=0.38]{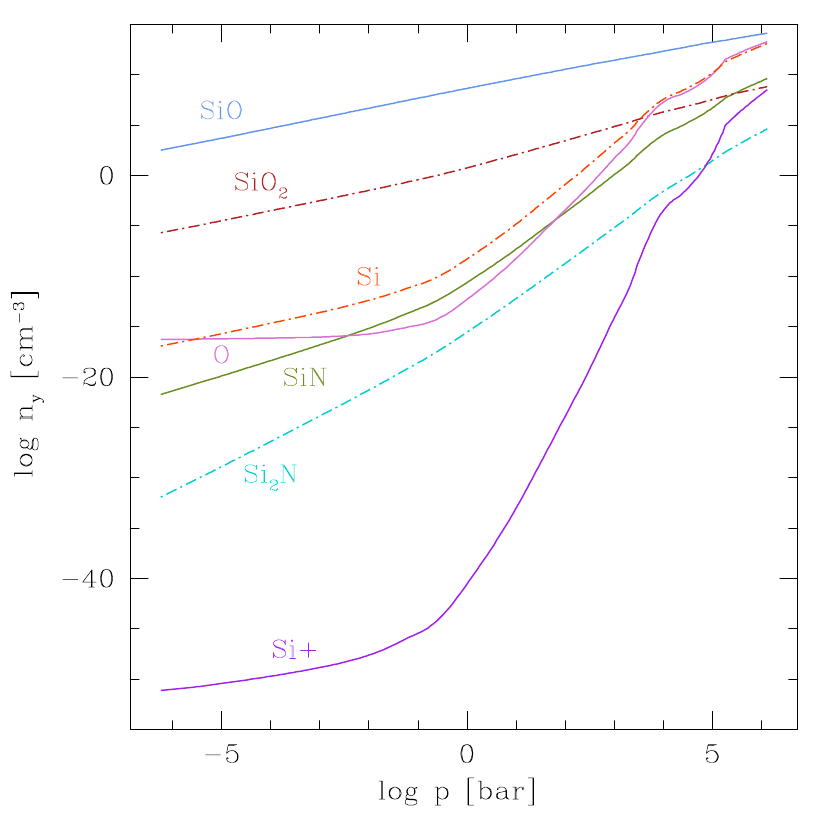}
 \caption{Comparison of number densities (cm$^{-3}$) for different
   Ti-binding (top) and Si-binding (buttom) molecules for a {\sc Drift-Phoenix}  (T$_{\rm
  gas}$, p$_{\rm gas}$) structure for T$_{\rm eff}$=1600K,
log(g)=3.0 and solar metallicity.}
\label{fig:TiSimol}
\end{figure}

\subsubsection{Chemical equilibrium routine}\label{sec:equilibrium-chem}
A combination of $155$ gas-phase molecules (including $33$ complex
carbon-bearing molecules), $16$ atoms, and various ionic species were
used under the assumption of local thermodynamic equilibrium
(LTE). For more details, please refer to Bilger, Rimmer \& Helling
(2013), and for the thermodynamical data used Helling, Winters \&
Sedlmayr (2000). The Grevesse, Asplund \& Sauval (2007) solar
composition is used for calculating the gas-phase chemistry outside
the metal depleted cloud layers and before cloud formation. No solid
particles were included in the chemical equilibrium calculations but
their presence influences the gas phase by the reduced element
abundances due to cloud formation and the cloud opacity impact on the
radiation field, both accounted for in the {\sc Drift-phoenix} model
simulations. We utilize {\sc Drift-Phoenix} model atmosphere (T$_{\rm
  gas}$, p$_{\rm gas}$) structures as input for our calculations.

\subsection{Results for molecule and cluster abundances}\label{ssec:clusterabund}

As pressure and temperature increase inwards the atmosphere, the
abundance of all gas species increase in chemical equilibrium
(Figs.~\ref{fig:TiSimol}).  For comparison, both Ti and Si
combinations are shown because the number densities of TiO$_2$ and SiO
are input properties for Eqs.~\ref{eq:J*final},~\ref{eq:p(n)}.  SiO
has generally a higher number density than TiO$_{2}$ since the element
abundance of Si is considerable larger than that of Ti. This might
suggest SiO as a more suitable nucleation species than TiO$_2$ and we
will investigate this question in Sects.~\ref{ssec:SiOnuc}
and~\ref{sec:cloud}.  Figure~\ref{fig:TiSimol} (top) demonstrates that
TiO$_2$ is the most abundant Ti-binding gas-species in almost the
entire atmosphere followed by TiO and the Ti atom. TiO is more
abundant than TiO$_2$ in the high-temperature part of the atmospheric
structure. SiO is the most abundant Si-binding molecules followed by
SiO$_2$ and Si.

\begin{figure}
 \includegraphics[scale=0.285]{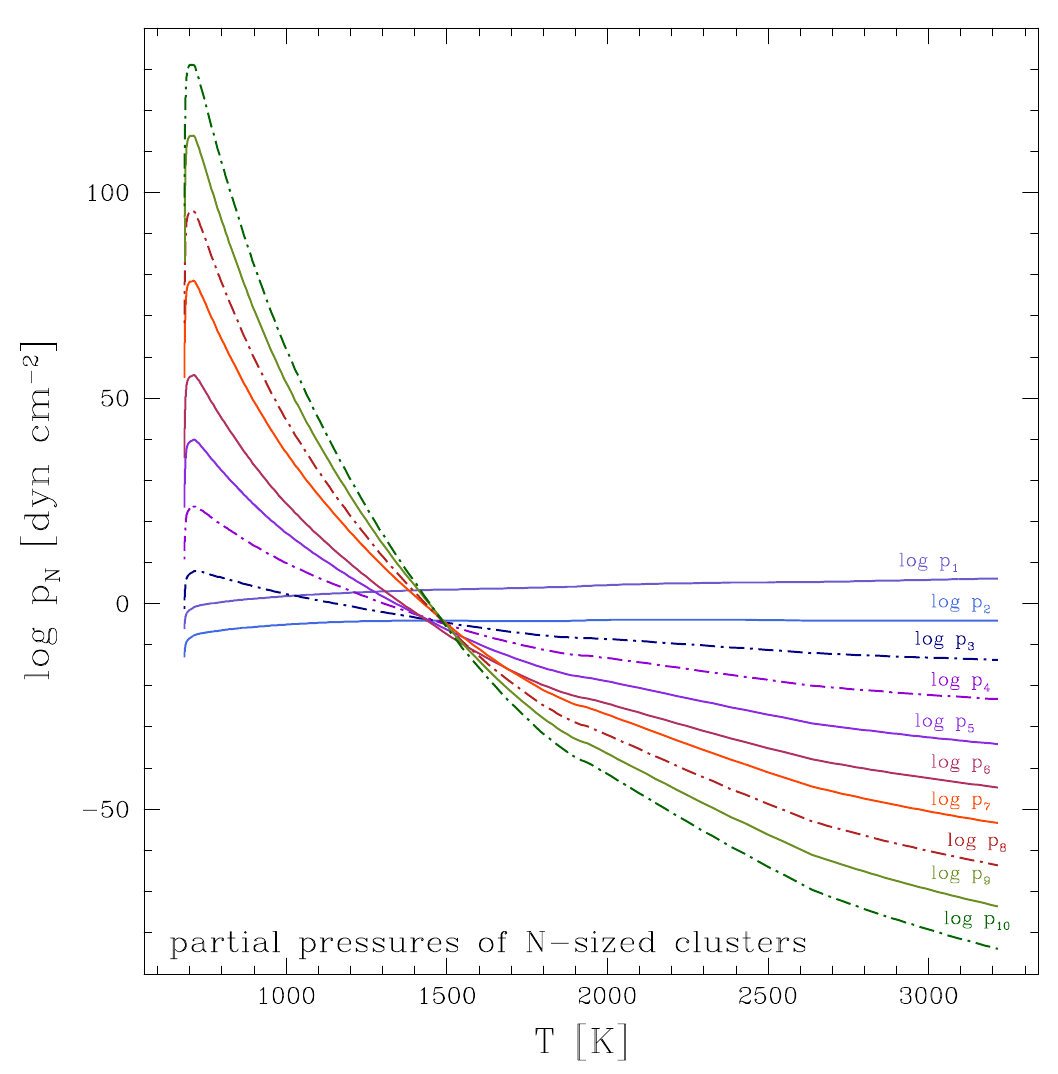}
 \caption{Partial pressures $\stackrel{\circ}{p}$(N) for N=1,10 (dyn
   cm$^{-2}$) in chemical equilibrium. Each calculation uses the
   standardized pressure p$^{\st}$ and the Gibbs free energies from Sect.~\ref{ThETiO2}.}
\label{fig:p(N)}
\end{figure}
As part of our assessment of the TiO$_2$ nucleation, we show the
partial pressure, $\stackrel{\circ}{p}(N)$ [dyn cm$^{-2}$]
(Eq.~\ref{eq:p(n)}), for each (TiO$_{2}$)$_N$-cluster in
Fig.~\ref{fig:p(N)}.  Both $\stackrel{\circ}{p}$(1) ($= n_{\rm TiO_2} k
T_{\rm gas}$) and $\stackrel{\circ}{p}$(2) maintain fairly constant
pressures. For $N>2$, the curves become more dynamic. They start at
higher and higher magnitudes, increase quickly and then drop off. The
order of the curves are also interesting, with the higher $N$ partial
pressures reaching higher values at lower temperature (lower gas
pressures). This findings support our expectation that bigger clusters
become more stable and more abundant with decreasing temperatures and
that they are unstable and of low abundance at high temperatures. 

\section{Seed formation rates}\label{sec:J*many}

Based on the results from the previous Sects.~\ref{ssec:SurfTen} and
~\ref{ssec:clusterabund}, we are now in the position to calculate and
compare seed formation rates (nucleation rates). We present our
updated results for TiO$_2$ as the nucleation species considered in
our previous works. We compare SiO-nucleation based on updated vapour
pressure data. Gail et al. (2013) recently suggested that
SiO-nucleation could be more efficient than TiO$_2$-nucleation. We
will test this hypothesis here and as part of our cloud formation
model in Sect.~\ref{sec:cloud}.

\subsection{TiO$_{2}$ nucleation rate}

\begin{figure}
 \includegraphics[scale=0.28]{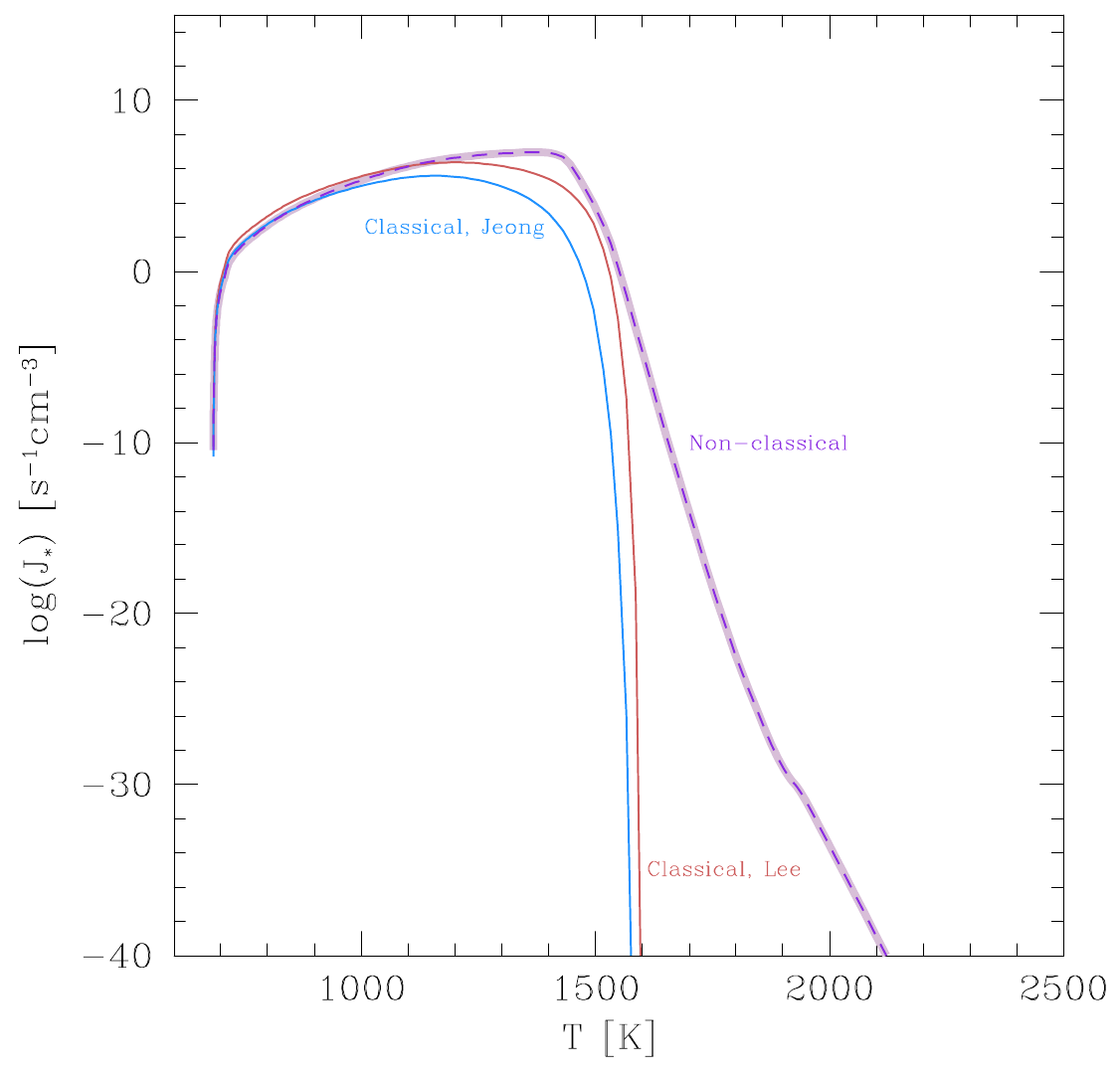}
 \caption{Nucleation rates J$_*$ with three methods: classical with
   $\sigma_{\infty}$ = 618 erg cm$^{-2}$, Jeong et al (2000);
   classical with temperature dependent $\sigma_{\infty}$; and
   non-classical based on Gibbs free energies for TiO$_2$.}
\label{fig:pureJ*}
\end{figure}

\paragraph{Results for classical nucleation theory:}

Surface tension values have a direct impact on the nucleation rate in
the classical nucleation theory approach (Sect.~\ref{ssec:clnucT}).
In order to assess this impact, the new TiO$_2$ surface tension was
tested in our nucleation routines and nucleation rates calculated for
a given ($T_{\rm gas}$, $p_{\rm gas}$) model atmosphere profile.
Figure~\ref{fig:pureJ*} demonstrates that the difference in nucleation
rate, $J_*$, from our new data to the value from Jeong et al. (2000)
is not very significant.

\begin{figure}
\centering
 \includegraphics[scale=0.35]{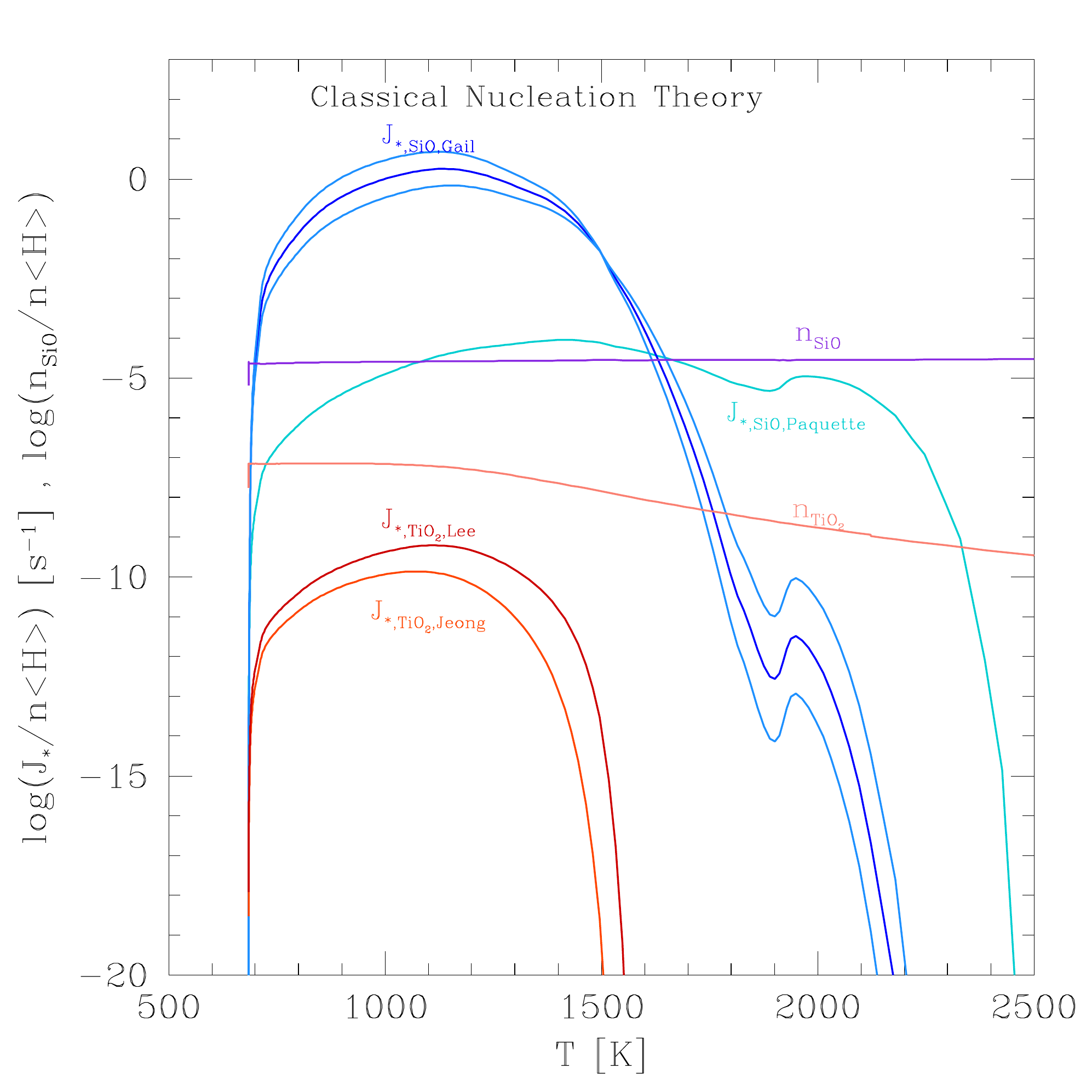}
 \caption{Number densities (cm$^{-3}$) and nucleation rates J$_*$
   (s$^{-1}$cm$^{-3}$) for both TiO$_{2}$ and SiO. Number densities
   calculated from the {\sc Drift-Phoenix} model (Ch. Helling et al
   2006). J$_{\rm *, Jeong}$ is the classical nucleation rate
   calculated with $\sigma$ = 618 erg cm$^{2}$, J$_{\rm *, Lee}$ is
   the with temperature dependent $\sigma$. J$_{\rm *,Paquette}$ and
   J$_{\rm *, Gail}$ have both been calculated using new vapour
   pressures (Wetzel et al. 2012). Blue lines surrounding J$_{\rm *,
     Gail}$ are the upper and lower boundaries. These nucleation rates
   were calculated for an undepleted gas-phase.}
\label{fig:JSiO}
\end{figure}

\paragraph{Results for non-classical nucleation theory:}

Converting all partial pressures, $\stackrel{\circ}{p}(N)$, for all
(TiO$_{2}$)$_N$-cluster into number densities allows us to use the
Becker-D{\"o}ring method in order to calculate the nucleation rate
J$_*$ (Gail \& Seldmayr 2014). This is different from the classical
nucleation rate in that we use the Gibbs free energies of formation
for each individual cluster, without the need to derive a surface
tension.

Figure~\ref{fig:pureJ*} shows that at the lowest temperatures, the non-classical nucleation rate increases
quickly with the temperature until 700K where the rate increases more
slowly, to around 1800K, and then drops (though not as quickly as the
classical curves).  At the higher temperatures, the molecules will
have sufficient energies that collisions breaking them apart will
happen as often as they coalesce.  Though the non-classical values are
visibly different from the classical, it is similar in magnitude
to the classical data with a temperature varying $\sigma$,
particularly in the 700-1500K region of the model atmosphere considered here.

\subsection{SiO nucleation}\label{ssec:SiOnuc}
Stimulated by the recent paper by Gail et al. (2013), we compare the
nucleation rate of SiO to our TiO$_2$-values from the previous
sections. Since the number density of SiO is much greater than
TiO$_{2}$, it is not ridiculous to expect that the nucleation rate for SiO would
also be larger than that of TiO$_{2}$.  Gail et al (2013) provide the
following analytic expression for the SiO nucleation rate including the updated
vapour pressure from Wetzel et al. (2012),
\begin{equation}
\label{eq:JSiO}
J_*^{\rm SiO} = n^2_1 \exp{\left((1.33\pm3.1)-\frac{(4.40\pm0.61)\cdot10^{12}}{T^3(\ln S)^2}\right)},
\end{equation}
where $n_1=\stackrel{\circ}{f}$(1) and all other variable have the
same meaning as before. Calculating $J_*^{\rm SiO}$ for the same model
atmosphere structure as before, we find that SiO nucleates at a much
higher rate compared to our TiO$_2$ results. We also demonstrate in
Fig.~\ref{fig:JSiO} the changes in the SiO-nucleation rates alone
through the update in vapour pressure data, $J_{\rm *, Paguette}$ vs.
$J_{\rm *, Gail}$.  There are similarities between the two SiO rates,
the double peaks occur at approximately the same temperatures,
indicating that both methods create similar effects at these
temperatures. These differences resulting from updated vapour pressure
data can not account for the differences between the SiO and the
TiO$_2$ nucleation rates.

\begin{figure}
\centering
 \includegraphics[scale=0.5]{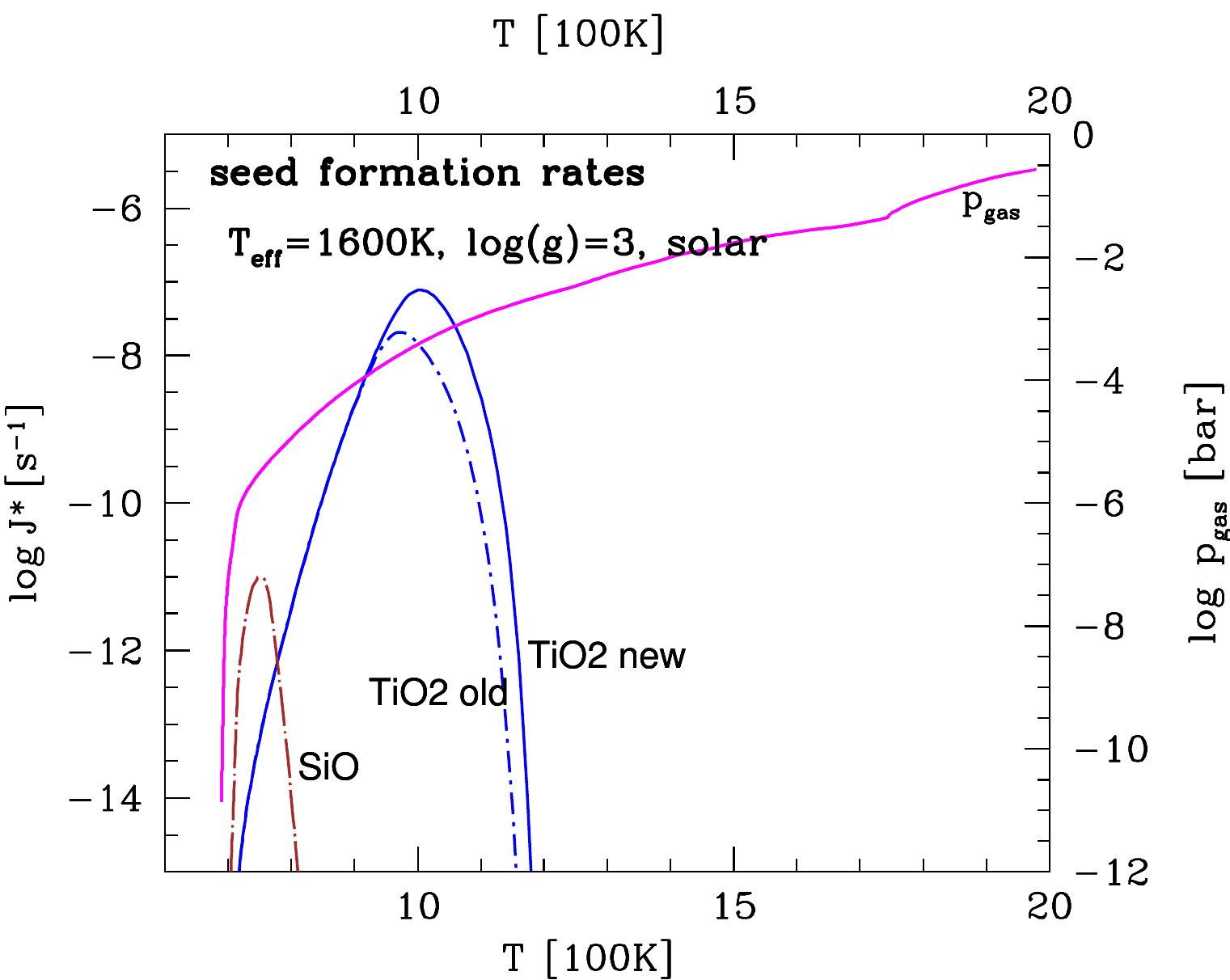}\\
 \includegraphics[scale=0.5]{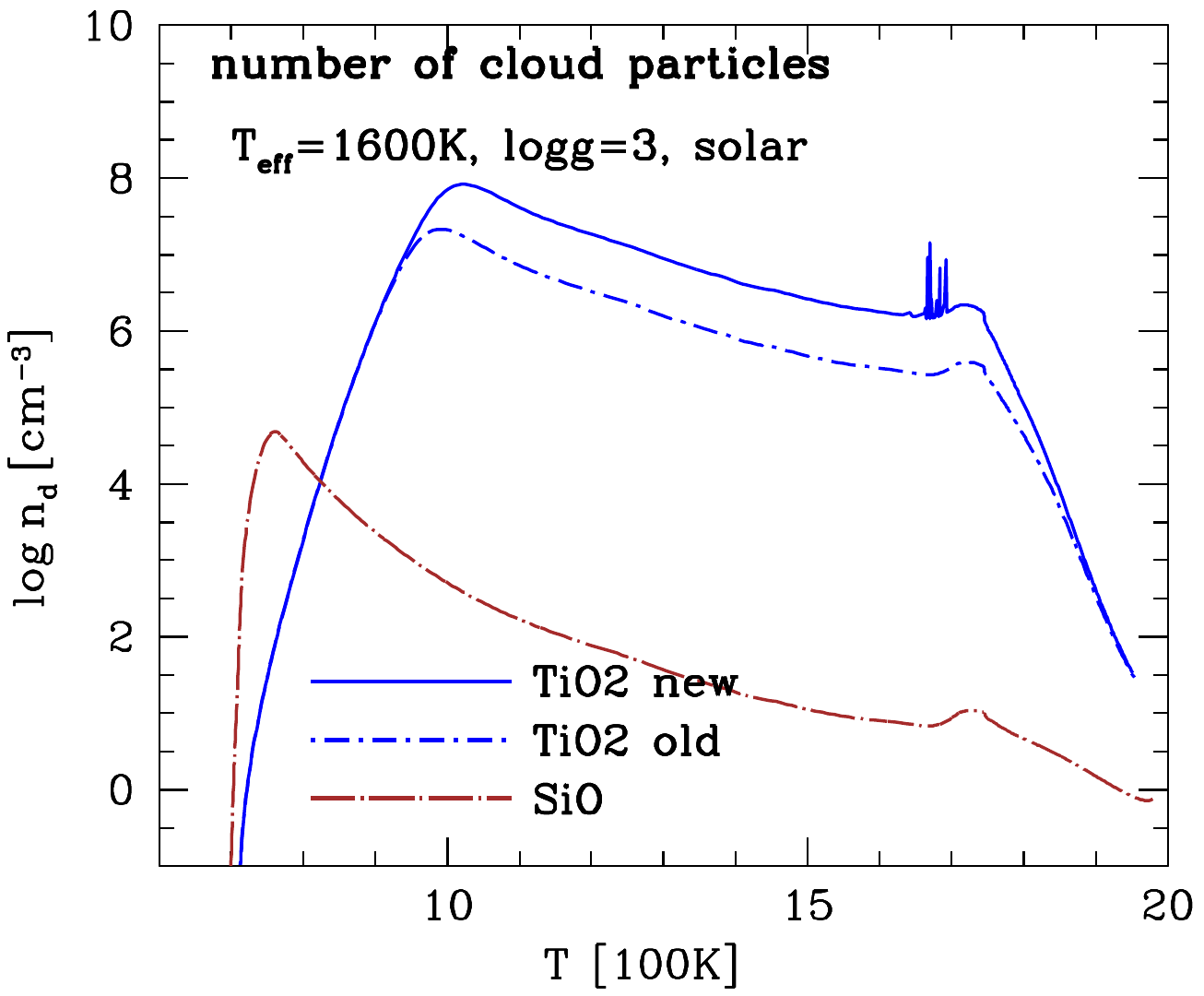}\\
 \includegraphics[scale=0.5]{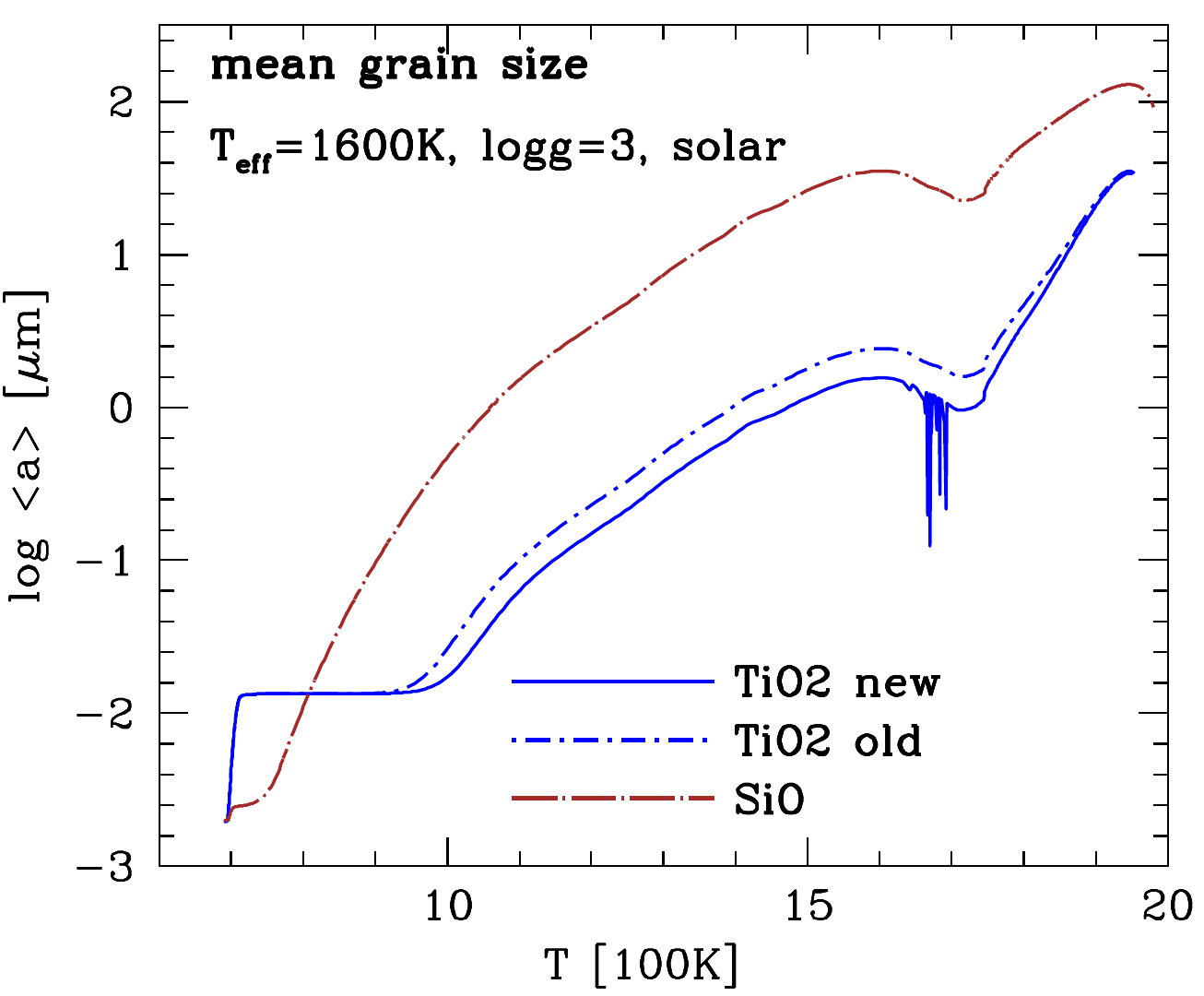}
 \caption{TiO$_2$- and SiO-nucleation rates (top) calculated as part of
   the cloud formation model and their effect on the number density of
   cloud particles (middle) and the mean grain size (bottom). The
   calculations include nucleation, growth/evaporation, element
   conservation, gravitational settling and convective
   replenishment. The same {\sc Drift-Phoenix} model structure for
   T$_{\rm eff}=1600$K, log(g)=3.0, and solar metallicity as in
   Fig.\ref{fig:pureJ*} was utilised.}
\label{fig:Jnda}
\end{figure}

\section{Impact on cloud formation}\label{sec:cloud}

Cloud formation in brown dwarfs and giant gas planets needs to start
with the formation of condensation seeds in contrasts to Earth where
weather cloud formation is started through the injection of seed
particles (e.g. volcano eruptions or sand storms) into the
atmosphere. Jeong et al. (1999) demonstrate that it is not obvious
which species would be the best choice for a nucleation species as
part of a dust / cloud particle formation model. Gail et al. (2013)
and Helling \& Fomins (2013) further argue that the complex silicate
seeds (e.g. Mg$_2$SiO$_4$[s], Al$_2$O$_3$[s]) can only form from
molecules that are available in the gas-phase. The SiO and
  TiO$_2$ are available in abundance in the gas phase
  (Fig.~\ref{fig:TiSimol}) but Mg$_2$SiO$_4$ does not exist as molecule,
  and Al$_2$O$_3$ is extremely low abundant (e.g. Fig. 5 in Helling \&
  Woitke 2006). Other Mg or Al binding molecules are abundant pointing
  to the possibility of heterogeneous nucleation. Hence, the formation
  of seed particles does not need to proceed via a homomolecular
homogeneous nucleation, but may well be formed by heteromolecular
homogeneous nucleation (e.g. Goumans \& Bromley 2013, Plane
2013).  Due to the lack of cluster data for more complex nucleation
  paths, we consider homomolecular homogeneous nucleation
  only.

 In principle, the condensing material does not care which seed
  particle is available as long as there is a surface to condense
  on. The need to identify the first condensate, or the most efficient
  nucleation species, arises if a model is built in order to study
  dust forming systems like, for example, clouds in brown dwarfs and
  exoplanets or dust in circumstellar shells. The two best
candidates with respect to stability and abundance in the gas phase
are TiO$_2$ and SiO. We are now in the position to test how the new
material data for TiO$_2$ (Sect.~\ref{ThETiO2}) and the updated
saturation vapour pressure for SiO (Eq.~\ref{eq:JSiO}) affects our
cloud formation results. Our results in this paper so far lead us to
expect only moderate differences from the updated TiO$_2$-nucleation
rate (Fig.~\ref{fig:pureJ*}), but substantial differences if
considering SiO instead of TiO$_2$ as nucleation
species. Figure~\ref{fig:JSiO} suggests a considerably more efficient
SiO seed formation compared to TiO$_2$ seed formation.  In this
section, we will demonstrated that it is misleading to consider seed
formation as a single process only.  The nucleation process needs to
be considered in combination with other, element consuming cloud/dust
formation processes in order to reliably approach the question about
the best suitable condensation seed species.

\subsection{Approach}
We assess the impact of the nucleation description that is part of our
cloud formation model on the resulting cloud structure details. Our
cloud formation model describes the formation of clouds by nucleation,
subsequent growth by chemical surface reactions on-top of the seeds,
evaporation, gravitational settling, element conservation and convective
replenishment (Woitke \& Helling 2003, 2004; Helling \& Woitke 2006,
Helling et al. 2008a). The effect of nucleation, growth \& evaporation on the remaining elements in the gas phase is fully accounted for (Eqs. 10 in Helling, Woitke \& Thi 2008a).  The surface growth causes the grains to grow to
$\mu$m-sized particles of a mixed composition of those solids taken
into account. For this study, we consider 12 growth species that grow
by 60 gas-solid surface reactions (Helling et al. 2008a). We use the
same {\sf Drift-Phoenix} model atmosphere as described in
Sect.~\ref{ss:DF} as input for our more complex cloud formation code.
 
\subsection{Results}
\begin{figure*}
 \includegraphics[scale=0.7]{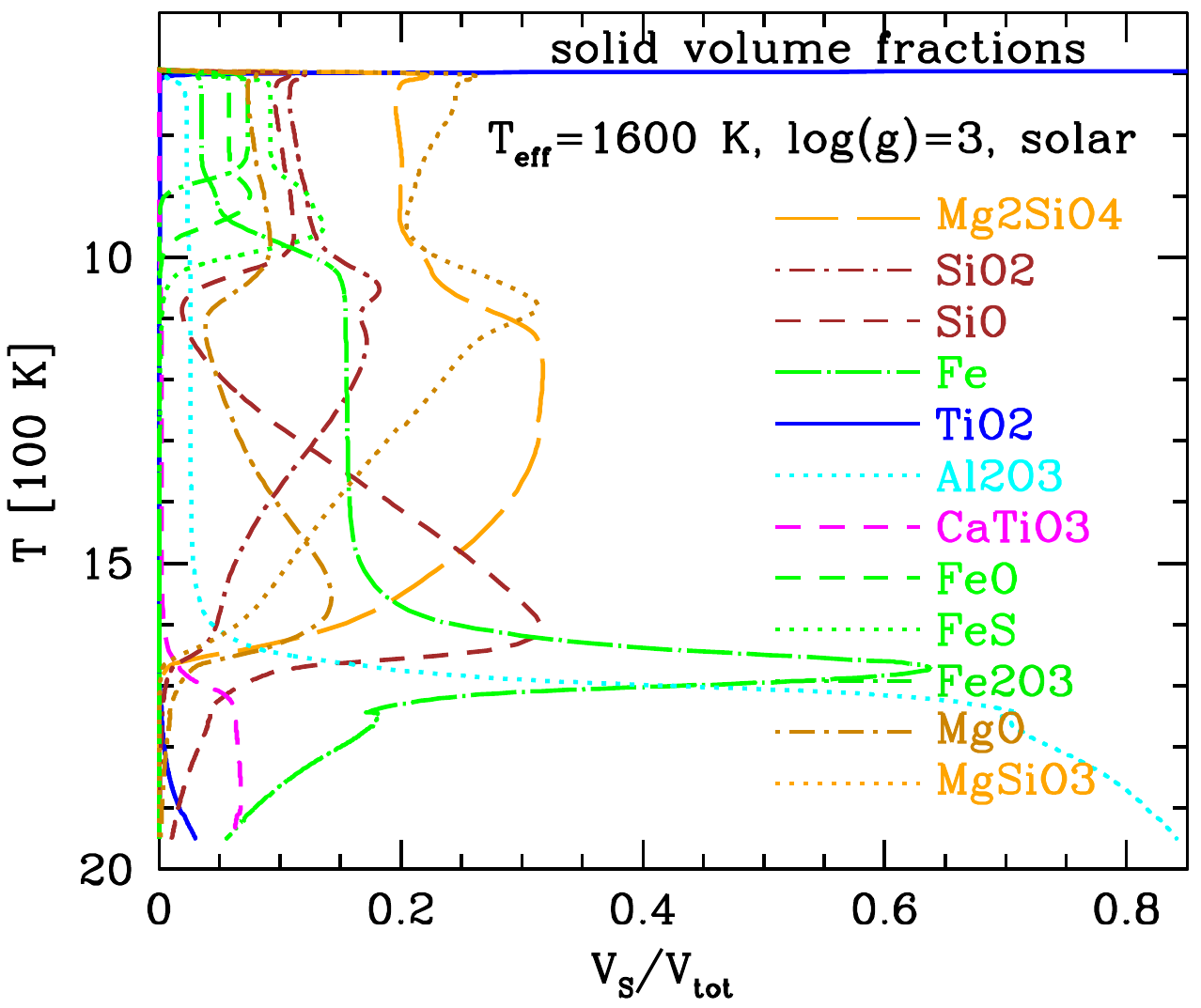}
 \includegraphics[scale=0.7]{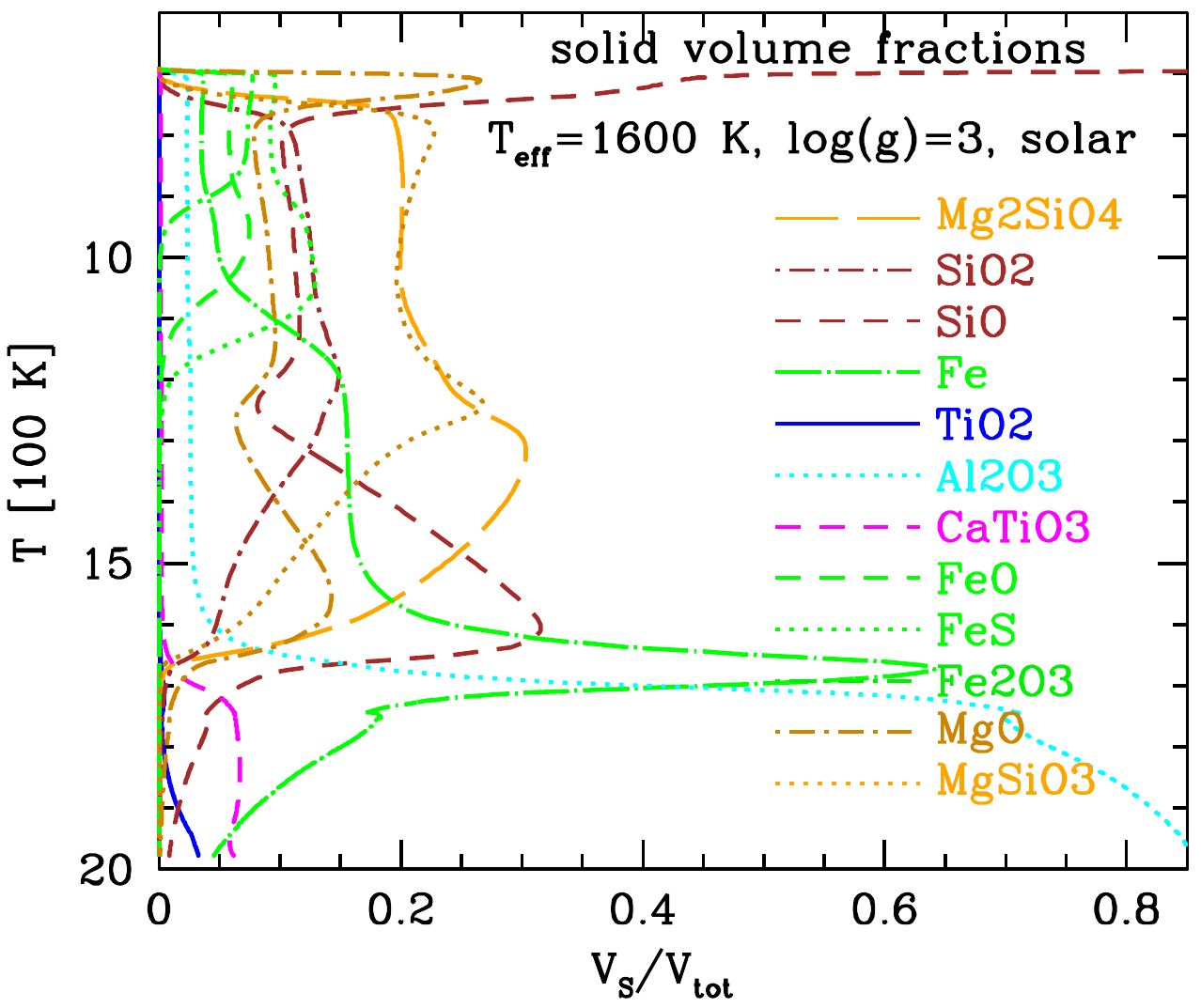}
 \caption{The differences in the cloud particle material compositions
   for TiO$_2$-nucleation (left) and SiO (right). The nucleation
   species (TiO$_2$ or SiO molecule) are also surface growth species
   (as in Table~1 in Helling, Woitke \& Thi 2008). The main
   differences occur in the uppermost part of the cloud, the haze
   layer: SiO[s]/MgO[s] with impurities of FeS[s], FeO[s],
   Fe$_2$O$_3$[s], and Al$_2$O$_3$[s] in the case of SiO-nucleation (right),
   and MgSiO$_3$[s]/Mg$_2$SiO$_4$[s] with impurities of all other
   solids plus a very thin TiO$_2$[s] layer at the very top in the
   case of TiO$_2$[s]-nucleation (left).}
\label{fig:Vs}
\end{figure*}

Figure~\ref{fig:Jnda} demonstrates the impact of the nucleation
treatment on the cloud formation processes and the resulting cloud
properties. Most importantly, if considered as part of an interacting
set of processes, the TiO$_2$-seed formation is more efficient than
the SiO-seed formation (top panel with nucleation rates) which
deviates from our privious expectation triggered by
Fig.~\ref{fig:JSiO}. The reason is that the elements Si and O are part
of many silicate materials (SiO$_2$[s],MgSiO$_3$[s], Mg$_2$SiO$_4$[s],
$\ldots$) that are already thermally stable and therefore grow
efficiently as soon as the seed particles emerge from the gas
phase. Ti-binding growth species are much less abundant due to the low
Ti element abundances to start with (Fig.~\ref{fig:TiSimol}).  Hence,
an assessment of the importance of a seed forming species always need
to be performed in connection with the growth process, else it leads
to wrong conclusions regarding the best suited nucleation species.  As
a consequence of SiO being a very inefficient nucleation species, less
cloud particles form. Figure~\ref{fig:Jnda} (middle) show that a
SiO-seeded cloud would have $>10^3$ times less cloud particles with an
increasing difference for increasing atmospheric depth. Instead the
material is consumed by growth leading to grains up to a size of
100$\mu$m at the inner cloud edge.

Figure~\ref{fig:Vs} demonstrates that the overall mean material
composition of the mineral cloud does not change significantly between
TiO$_2$-seeded and SiO-seeded clouds. However, the upper-most part
which is often referred to as haze-layer has a fundamentally different
composition depending on the condensation seed species considered:
SiO[s]/MgO[s] with impurities of FeS[s], FeO[s], Fe$_2$O$_3$[s], and
Al$_2$O$_3$[s] in the case of SiO-nucleation, and
MgSiO$_3$[s]/Mg$_2$SiO$_4$[s] with impurities of all other solids plus
a very thin TiO$_2$[s] layer at the very top in the case of
TiO$_2$[s]-nucleation.

\section{Summary}
The formation of condensation seeds is the initial step to cloud
formation in astrophysical objects without a crust, hence, for objects
like Super Nova, AGB-stars, M-dwarfs, brown dwarfs, and giant gas
planets. The long standing question is if it is possible to identify a
first condensate that kicks off the whole condensation process. This
question has long been debated and high-temperature condensates like
solid iron seeds forming from the gas phase from (Fe)$_N$-clusters
(John \& Sedlmayr 1997) or MgO seeds forming from (MgO)$_N$-clusters
(K\"ohler \& Sedlmayr 1997) needed to be dismissed because of only
small clusters being thermodynamically stable or not very abundant.
TiO$_2$ seed formation is attractive because of the stability of the
(TiO$_2$)$_N$-clusters and their relative abundance. The same
arguments are made for SiO but despite SiO's higher abundance compared
to TiO$_2$, it's nucleation rate did fall short of TiO$_2$ (Jeong et
al. 2000). Gail et al. (2013) re-consider SiO nucleation for AGB stars
and suggest it to be a favorable seed formation species based on new
vapour data. Based on updated (TiO$_2$)$_N$-clusters we investigate
under which conditions this finding could be relevant for substellar
atmospheres.

In this paper, we have presented updated Gibbs free energies of
TiO$_{2}$-clusters using computational chemistry for newly available
molecule geometries (Calatayud et al 2008, Syzgantseva et al 2010). The more stable cluster geometries compared
to Jeong et al. (2000) from chemistry literature yielded a temperature
dependent surface tension with an average value of $\sigma_{\infty}$ =
480.6 erg cm$^{-2}$ when fitted with the modified nucleation theory
model.  This new surface tension was then used in conjunction with
chemical abundance routines to calculate a nucleation rate for various
temperatures and TiO$_{2}$ number densities for an example atmosphere
representative for a young brown dwarfs or a giant gas plant.  The new
value approximately doubles the rate of nucleation for the specie.
The non-classical TiO$_2$ nucleation rate was calculated using the
newly calculated Gibbs free energies which obtained higher results
than those obtained through classical means.  Inspired by newly available vapour pressure data, we show that
SiO nucleation can only be more efficient than TiO$_2$ nucleation if
no other element depletion processes are taking place. Hence, TiO$_2$
remains the more efficient nucleation species of the two as nucleation
and surface growth will take place simultaneously as both processes
require a supersaturated gas.

 \begin{acknowledgements}
{We highlight financial support of the European Community under the
  FP7 by an ERC starting grant.  A scholarship fund for GL and HG
  provided by the Royal Astronomical Society is highly acknowledged.
  We thank Peter Woitke for general discussions
  on the paper's topic.  The computer support at the School of Physics
  \& Astronomy in St Andrews is gratefully acknowledged.}
 \end{acknowledgements}



\appendix
\section{Energy Tables}
References to \textbf{'Jeong's Geometry'} or similar refer to the original geometries as found in Jeong et al (2000). References to \textbf{'Bromley's Geometry'} or similar 
refer to the current (2012) most stable (TiO$_{2}$)$_{N}$ cluster geometries.

\begin{table*}
\begin{tabular}{|c|c|c|c|c|c||c|} \hline
 Molecule		& Sym.	& $E_{0}$[Hartree]	& $E_{\rm zpe}$[eV] & $D_{at}$[eV]	& $D_{0}$[eV] & $D^{exp}_{0}$ [eV] \\ \hline
 Ti($^{3}$F)		& -		& -849.299325			& 0.00			& -		& -		& -			\\ 
 O($^{3}$P)		& - 		& -75.060623			& 0.00			& -		& -		& -			\\ \hline
$ TiO_{2}$		& $C_{2\nu}$	& -999.963285			& 0.15			& 12.57		& 12.42		& 13.06			\\
$ (TiO_{2})_{2}$	& $C_{2h}$	& -2000.124814			& 0.40			& 30.64		& 30.24		& 31.27 $\pm$ 0.68	\\
$ (TiO_{2})_{3}	$	& $C_{2}$	& -3000.269292			& 0.63			& 48.22		& 47.59		&			\\
$ (TiO_{2})_{3\ Bromley}$& $C_{s}$	& -3000.287915			& 0.63			& 48.73		& 48.10		&			\\
$ (TiO_{2})_{4}$	& $C_{2h}$	& -4000.417659			& 0.86			& 65.91		& 65.05		&			\\
$ (TiO_{2})_{4\ Bromley}$& $C_{2\nu}$	& -4000.463105			& 0.87			& 67.15		& 66.28		&			\\
$ (TiO_{2})_{5}	$	& $C_{2}$	& -5000.564391			& 1.09			& 83.55		& 82.46		&			\\
$ (TiO_{2})_{5\ Bromley}$& $C_{s}$	& -5000.632153			& 1.11			& 85.41		& 84.30		&			\\
$ (TiO_{2})_{6}	$	& $C_{2h}$	& -6000.711386			& 1.32			& 101.20	& 99.88		&			\\
$ (TiO_{2})_{6\ Bromley}$& $C_{2}$	& -6000.837167			& 1.34			& 104.64	& 103.30	&			\\
$ (TiO_{2})_{7\ Bromley}	$& $C_{s}$	& -7001.011900			& 1.58			& 123.06	& 121.48	&			\\
$ (TiO_{2})_{8\ Bromley}	$& $C_{2h}$	& -8001.195277			& 1.82			& 141.71	& 139.89	&			\\
$ (TiO_{2})_{9\ Bromley}$& $C_{s}$	& -9001.372943			& 2.05			& 160.19	& 158.14	&			\\
$ (TiO_{2})_{10\ Bromley}$& $C_{1}$	& -10001.567429			& 2.27			& 179.13	& 176.86	&			\\ \hline
 \end{tabular}
 \caption{DFT/B3LYP energies for Ti, O and $(TiO_{2})_{n}$. $E_{0}$ is the internal energy. $E_{\rm zpe}$ is the zero-point energy. $D_{at}$ atomization energy.$D_{0}$ atomization energy minus $E_{\rm zpe}$ and 
 $D^{\rm exp}_{0}$ the expected experimental results.}
\end{table*}

\begin{table*}
 \begin{tabular}{|c|c|c|c|c|} \hline
 Molecule			& $D_{0}$(n) [eV] 	& $D_{0}$(n)/n [eV]	& $\Delta$E(n) [eV] 	& $D_{0}$(n)/(x + y) [eV] \\ \hline
$ TiO_{2}$			& 12.42			& 12.42			& 12.42				& 4.14	\\
$ (TiO_{2})_{2}$		& 30.24			& 15.12			& 18.00				& 5.04	\\
$ (TiO_{2})_{3}	$		& 47.59			& 15.86			& 17.35				& 5.29	\\
$ (TiO_{2})_{3\ Bromley}$	& 48.10			& 16.03			& 17.86				& 5.34	\\
$ (TiO_{2})_{4}$		& 65.05			& 16.26			& 17.46				& 5.42	\\
$ (TiO_{2})_{4\ Bromley}$	& 66.28			& 16.57			& 18.18				& 5.52	\\
$ (TiO_{2})_{5}	$		& 82.46			& 16.49			& 17.41				& 5.50	\\
$ (TiO_{2})_{5\ Bromley}$	& 84.30			& 16.86			& 18.02				& 5.62	\\
$ (TiO_{2})_{6}	$		& 99.88			& 16.65			& 17.42				& 5.55	\\
$ (TiO_{2})_{6\ Bromley}$	& 103.30		& 17.22			& 19.00				& 5.74	\\
$ (TiO_{2})_{7\ Bromley}	$	& 121.48		& 17.35			& 18.18				& 5.78	\\
$ (TiO_{2})_{8\ Bromley}	$	& 139.89		& 17.49			& 18.41				& 5.83	\\
$ (TiO_{2})_{9\ Bromley}$	& 158.14		& 17.57			& 18.25				& 5.86	\\
$ (TiO_{2})_{10\ Bromley}$	& 176.86		& 17.69			& 18.72				& 5.90	\\ \hline
 \end{tabular}
 \caption{Binding energies for $(TiO_{2})_{n}$ clusters. Total binding energy $D_{0}$(n), binding per monomer $D_{0}$(n)/n, energy gain $\Delta$E(n) by adding a monomer and the binding energy per atom 
 $D_{0}$(n)/(x + y). Bromley energies have larger binding energies that the original Jeong geometries.}
\end{table*}
\begin{table}[thb]
\begin{tabular}{|c|lcr|}	\hline 
Element & \(\Delta_{f}H^{\circ}(X,0K)\) & \(H^{\circ}(298K) - H^{\circ}(0K)\) & \(S^{\circ}(298K)\) \\ \hline
Ti	& 112.5525812				 & 1.154397705 				& 7.351578203 	\\
O	& 58.98422559				 & 1.037643403 				& 24.51565487 	\\ \hline
\end{tabular}
 \caption{Chemical information on the atoms Ti and O. 
 Note: These values are in \(kcal\ mol^{-1}\), \(kcal\ mol^{-1}\) and \(cal\ K^{-1} mol^{-1}\) respectively.
 JANAF thermochemical tables (O 1982, Ti 1979)}
\end{table}
\clearpage

\section{Thermochemical Tables of (TiO$_{2}$)$_{N}$}\label{appB}

\begin{table*}
\begin{center}
\caption{ The calculated thermochemical values of \(TiO_{2}\)}
 \begin{tabular}{|r|r|r|r|c|c|}		\hline
 T	& \(S^{\circ}\) 	& \([-G^{\circ}-H^{\circ}(T_{r})]/T \)	& \(H^{\circ} - H^{\circ}(T_{r})\)	& \(\Delta_{f}H^{\circ}	\) 	& \(\Delta_{f}G^{\circ}\) \\
 K	& \(J mol^{-1}K^{-1}\)	& \(J mol^{-1}K^{-1}\)			& \(kJ mol^{-1}\)			& \(kJ mol^{-1}\)			& \(kJ mol^{-1}\)		\\ \hline
        0   &      0.000   &  Infinite   &   -10.951   &  -233.980   &     -233.980   \\
      100   &   216.499   &   290.042   &    -7.606   &  -234.100   &  -237.593   \\
      200   &   241.559   &   259.146   &    -3.967   &  -235.289   &  -240.659   \\
      298.15   &   257.636   &   255.419   &    0.000   &  -236.542   &  -243.021   \\
      300   &   257.903   &   255.423   &     0.079   &  -236.564   &  -243.062   \\
      400   &   270.656   &   257.122   &     4.521   &  -237.706   &  -245.051   \\
\hline
 \end{tabular}
 \end{center}
\end{table*}

\clearpage
\begin{table*}
\begin{center}
\caption{ The calculated thermochemical values of \((TiO_{2})_{2}\)}
 \begin{tabular}{|r|r|r|r|c|c|}		\hline
 T	& \(S^{\circ}\) 	& \([-G^{\circ}-H^{\circ}(T_{r})]/T \)	& \(H^{\circ} - H^{\circ}(T_{r})\)	& \(\Delta_{f}H^{\circ}	\) 	& \(\Delta_{f}G^{\circ}\) \\
 K	& \(J mol^{-1}K^{-1}\)	& \(J mol^{-1}K^{-1}\)			& \(kJ mol^{-1}\)			& \(kJ mol^{-1}\)			& \(kJ mol^{-1}\)		\\ \hline
        0   &       0.000   &  Infinite   &   -18.885   &  -988.450   &    -988.450	\\
      100   &   262.813   &   407.720   &   -14.845   &  -991.339   &  -981.307   \\
      200   &   306.868   &   345.640   &    -8.354   &  -994.505   &  -969.994   \\
      298.15   &   340.624   &   337.693   &    0.000   &  -996.591   &  -957.477   \\
      300   &   341.201   &   337.693   &     0.173   &  -996.619   &  -957.233   \\
      400   &   369.762   &   341.476   &    10.129   &  -997.832   &  -943.901   \\
\hline
 \end{tabular}
 \end{center}
\end{table*}
\clearpage
\begin{table*}
\begin{center}
\caption{ The calculated thermochemical values of \textbf{Jeong's \((TiO_{2})_{3}\)}}
 \begin{tabular}{|r|r|r|r|c|c|}		\hline
 T	& \(S^{\circ}\) 	& \([-G^{\circ}-H^{\circ}(T_{r})]/T \)	& \(H^{\circ} - H^{\circ}(T_{r})\)	& \(\Delta_{f}H^{\circ}	\) 	& \(\Delta_{f}G^{\circ}\) \\
 K	& \(J mol^{-1}K^{-1}\)	& \(J mol^{-1}K^{-1}\)			& \(kJ mol^{-1}\)			& \(kJ mol^{-1}\)			& \(kJ mol^{-1}\)		\\ \hline
        0   &       0.000   &  Infinite   &   -28.116   & -1698.152   &       -1698.152   \\
      100   &   302.773   &   526.126   &   -22.792   & -1703.223   & -1679.030   \\
      200   &   369.250   &   430.378   &   -12.973   & -1707.887   & -1652.910   \\
      298.15   &   421.656   &   418.028   &    0.000   & -1710.575   & -1625.285   \\
      300   &   422.539   &   418.019   &     0.268   & -1710.610   & -1624.752   \\
      400   &   467.293   &   423.942   &    15.869   & -1711.761   & -1595.926   \\
\hline
 \end{tabular}
 \end{center}
\end{table*}
\clearpage
\begin{table*}
\begin{center}
\caption{ The calculated thermochemical values of \textbf{Bromley's \((TiO_{2})_{3}\)}}
 \begin{tabular}{|r|r|r|r|c|c|}		\hline
 T	& \(S^{\circ}\) 	& \([-G^{\circ}-H^{\circ}(T_{r})]/T \)	& \(H^{\circ} - H^{\circ}(T_{r})\)	& \(\Delta_{f}H^{\circ}	\) 	& \(\Delta_{f}G^{\circ}\) \\
 K	& \(J mol^{-1}K^{-1}\)	& \(J mol^{-1}K^{-1}\)			& \(kJ mol^{-1}\)			& \(kJ mol^{-1}\)			& \(kJ mol^{-1}\)		\\ \hline
        0   &      0.000   &  Infinite   &   -28.116   & -1747.136   &      -1747.136   \\
      100   &   302.668   &   526.022   &   -22.792   & -1752.207   & -1728.003   \\
      200   &   369.132   &   430.261   &   -12.973   & -1756.871   & -1701.871   \\
      298.15   &   421.551   &   417.923   &    0.000   & -1759.559   & -1674.238   \\
      300   &   422.434   &   417.914   &     0.268   & -1759.594   & -1673.705   \\
      400   &   467.181   &   423.832   &    15.869   & -1760.745   & -1644.865   \\
\hline
 \end{tabular}
 \end{center}
\end{table*}
\clearpage

\begin{table*}
\begin{center}
\caption{ The calculated thermochemical values of \textbf{Jeong's \((TiO_{2})_{4}\)}}
 \begin{tabular}{|r|r|r|r|c|c|}		\hline
 T	& \(S^{\circ}\) 	& \([-G^{\circ}-H^{\circ}(T_{r})]/T \)	& \(H^{\circ} - H^{\circ}(T_{r})\)	& \(\Delta_{f}H^{\circ}	\) 	& \(\Delta_{f}G^{\circ}\) \\
 K	& \(J mol^{-1}K^{-1}\)	& \(J mol^{-1}K^{-1}\)			& \(kJ mol^{-1}\)			& \(kJ mol^{-1}\)			& \(kJ mol^{-1}\)		\\ \hline
        0   &       0.000   &  Infinite   &   -37.516   & -2418.065   &      -2418.065   \\
      100   &   338.217   &   639.899   &   -30.724   & -2425.133   & -2386.327   \\
      200   &   427.169   &   510.600   &   -17.572   & -2431.294   & -2344.958   \\
      298.15   &   498.118   &   493.833   &    0.000   & -2434.601   & -2301.774   \\
      300   &   499.344   &   493.841   &     0.365   & -2434.640   & -2300.951   \\
      400   &   560.236   &   501.893   &    21.595   & -2435.747   & -2256.171   \\
\hline
 \end{tabular}
 \end{center}
\end{table*}
\clearpage

\begin{table*}
\begin{center}
\caption{ The calculated thermochemical values of \textbf{Bromley's \((TiO_{2})_{4}\)}}
 \begin{tabular}{|r|r|r|r|c|c|}		\hline
 T	& \(S^{\circ}\) 	& \([-G^{\circ}-H^{\circ}(T_{r})]/T \)	& \(H^{\circ} - H^{\circ}(T_{r})\)	& \(\Delta_{f}H^{\circ}	\) 	& \(\Delta_{f}G^{\circ}\) \\
 K	& \(J mol^{-1}K^{-1}\)	& \(J mol^{-1}K^{-1}\)			& \(kJ mol^{-1}\)			& \(kJ mol^{-1}\)			& \(kJ mol^{-1}\)		\\ \hline
        0   &      0.000   &  Infinite   &   -35.788   & -2537.384   &      -2537.384   \\
      100   &   308.759   &   607.077   &   -30.359   & -2545.814   & -2504.063   \\
      200   &   394.258   &   478.350   &   -17.649   & -2552.416   & -2459.498   \\
      298.15   &   465.492   &   461.487   &    0.000   & -2555.647   & -2413.092   \\
      300   &   466.726   &   461.496   &     0.368   & -2555.684   & -2412.209   \\
      400   &   528.034   &   469.604   &    21.742   & -2556.646   & -2364.189   \\
\hline
 \end{tabular}
 \end{center}
\end{table*}
\clearpage

\begin{table*}
\begin{center}
\caption{ The calculated thermochemical values of \textbf{Jeong's \((TiO_{2})_{5}\)}}
 \begin{tabular}{|r|r|r|r|c|c|}		\hline
 T	& \(S^{\circ}\) 	& \([-G^{\circ}-H^{\circ}(T_{r})]/T \)	& \(H^{\circ} - H^{\circ}(T_{r})\)	& \(\Delta_{f}H^{\circ}	\) 	& \(\Delta_{f}G^{\circ}\) \\
 K	& \(J mol^{-1}K^{-1}\)	& \(J mol^{-1}K^{-1}\)			& \(kJ mol^{-1}\)			& \(kJ mol^{-1}\)			& \(kJ mol^{-1}\)		\\ \hline
        0   &       0.000   &  Infinite   &   -47.010   & -3133.685   &       -3133.685   \\
      100   &   385.397   &   765.646   &   -38.689   & -3142.690   & -3090.446   \\
      200   &   496.994   &   602.678   &   -22.183   & -3150.324   & -3035.010   \\
      298.15   &   586.566   &   581.519   &    0.000   & -3154.241   & -2977.448   \\
      300   &   588.103   &   581.516   &     0.462   & -3154.284   & -2976.349   \\
      400   &   665.170   &   591.713   &    27.329   & -3155.337   & -2916.817   \\
\hline
 \end{tabular}
 \end{center}
\end{table*}

\clearpage
\begin{table*}
\begin{center}
\caption{ The calculated thermochemical values of \textbf{Bromley's \((TiO_{2})_{5}\)}}
 \begin{tabular}{|r|r|r|r|c|c|}		\hline
 T	& \(S^{\circ}\) 	& \([-G^{\circ}-H^{\circ}(T_{r})]/T \)	& \(H^{\circ} - H^{\circ}(T_{r})\)	& \(\Delta_{f}H^{\circ}	\) 	& \(\Delta_{f}G^{\circ}\) \\
 K	& \(J mol^{-1}K^{-1}\)	& \(J mol^{-1}K^{-1}\)			& \(kJ mol^{-1}\)			& \(kJ mol^{-1}\)			& \(kJ mol^{-1}\)		\\ \hline
        0   &      0.000   &  Infinite   &   -44.169   & -3311.595   &      -3311.595   \\
      100   &   327.321   &   701.588   &   -38.036   & -3322.786   & -3264.734   \\
      200   &   433.260   &   539.830   &   -22.251   & -3331.142   & -3203.082   \\
      298.15   &   523.066   &   518.565   &    0.000   & -3334.991   & -3139.266   \\
      300   &   524.619   &   518.569   &     0.465   & -3335.031   & -3138.051   \\
      400   &   602.139   &   528.820   &    27.492   & -3335.924   & -3072.192   \\
\hline
 \end{tabular}
 \end{center}
\end{table*}
\clearpage
\begin{table*}
\begin{center}
\caption{ The calculated thermochemical values of \textbf{Jeong's \((TiO_{2})_{6}\)}}
 \begin{tabular}{|r|r|r|r|c|c|}		\hline
 T	& \(S^{\circ}\) 	& \([-G^{\circ}-H^{\circ}(T_{r})]/T \)	& \(H^{\circ} - H^{\circ}(T_{r})\)	& \(\Delta_{f}H^{\circ}	\) 	& \(\Delta_{f}G^{\circ}\) \\
 K	& \(J mol^{-1}K^{-1}\)	& \(J mol^{-1}K^{-1}\)			& \(kJ mol^{-1}\)			& \(kJ mol^{-1}\)			& \(kJ mol^{-1}\)		\\ \hline
        0   &       0.000   &  Infinite   &   -56.566   & -3849.996   &      -3849.996   \\
      100  &   428.245   &   887.254   &   -46.671   & -3860.891   & -3794.774   \\
      200   &   562.553   &   690.552   &   -26.798   & -3869.986   & -3724.842   \\
      298.15   &   670.742   &   664.971   &    0.000   & -3874.508   & -3652.477   \\
      300   &   672.618   &   664.974   &     0.562   & -3874.552   & -3651.098   \\
      400   &   765.871   &   677.310   &    33.073   & -3875.544   & -3576.387   \\
\hline
 \end{tabular}
 \end{center}
\end{table*}
\clearpage
\begin{table*}
\begin{center}
\caption{ The calculated thermochemical values of \textbf{Bromley's \((TiO_{2})_{6}\)}}
 \begin{tabular}{|r|r|r|r|c|c|}		\hline
 T	& \(S^{\circ}\) 	& \([-G^{\circ}-H^{\circ}(T_{r})]/T \)	& \(H^{\circ} - H^{\circ}(T_{r})\)	& \(\Delta_{f}H^{\circ}	\) 	& \(\Delta_{f}G^{\circ}\) \\
 K	& \(J mol^{-1}K^{-1}\)	& \(J mol^{-1}K^{-1}\)			& \(kJ mol^{-1}\)			& \(kJ mol^{-1}\)			& \(kJ mol^{-1}\)		\\ \hline
        0   &       0.000   &  Infinite   &   -53.636   & -4180.234   &      -4180.234   \\
      100   &   346.986   &   805.262   &   -46.526   & -4193.914   & -4119.672   \\
      200   &   477.697   &   607.766   &   -27.069   & -4203.425   & -4041.310   \\
      298.15   &   586.971   &   581.920   &    0.000   & -4207.676   & -3960.669   \\
      300   &   588.865   &   581.932   &     0.564   & -4207.717   & -3959.137   \\
      400   &   682.794   &   594.355   &    33.312   & -4208.473   & -3876.085   \\
\hline
\end{tabular}
\end{center}
\end{table*}
\clearpage
\begin{table*}
\begin{center}
\caption{ The calculated thermochemical values of \textbf{Bromley's \((TiO_{2})_{7}\)}}
 \begin{tabular}{|r|r|r|r|c|c|}		\hline
 T	& \(S^{\circ}\) 	& \([-G^{\circ}-H^{\circ}(T_{r})]/T \)	& \(H^{\circ} - H^{\circ}(T_{r})\)	& \(\Delta_{f}H^{\circ}	\) 	& \(\Delta_{f}G^{\circ}\) \\
 K	& \(J mol^{-1}K^{-1}\)	& \(J mol^{-1}K^{-1}\)			& \(kJ mol^{-1}\)			& \(kJ mol^{-1}\)			& \(kJ mol^{-1}\)		\\ \hline
        0   &       0.000   &  Infinite   &   -61.739   & -4969.371   &       -4969.371   \\
      100   &   369.513   &   898.759   &   -53.705   & -4985.592   & -4895.446   \\
      200   &   518.536   &   670.197   &   -31.495   & -4997.179   & -4800.289   \\
      298.15   &   645.627   &   640.072   &    0.000   & -5002.223   & -4702.369   \\
      300   &   647.833   &   640.081   &     0.659   & -5002.271   & -4700.509   \\
      400   &   757.877   &   654.632   &    39.025   & -5002.993   & -4599.720   \\
\hline
\end{tabular}
\end{center}
\end{table*}
\clearpage
\begin{table*}
\begin{center}
\caption{ The calculated thermochemical values of \textbf{Bromley's \((TiO_{2})_{8}\)}}
 \begin{tabular}{|r|r|r|r|c|c|}		\hline
 T	& \(S^{\circ}\) 	& \([-G^{\circ}-H^{\circ}(T_{r})]/T \)	& \(H^{\circ} - H^{\circ}(T_{r})\)	& \(\Delta_{f}H^{\circ}	\) 	& \(\Delta_{f}G^{\circ}\) \\
 K	& \(J mol^{-1}K^{-1}\)	& \(J mol^{-1}K^{-1}\)			& \(kJ mol^{-1}\)			& \(kJ mol^{-1}\)			& \(kJ mol^{-1}\)		\\ \hline
        0   &       0.000   &  Infinite   &   -70.471   & -5781.203   &       -5781.203   \\
      100   &   385.397   &   992.932   &   -61.615   & -5800.067   & -5693.352   \\
      200   &   556.199   &   730.611   &   -36.151   & -5813.226   & -5580.927   \\
      298.15   &   702.091   &   696.050   &    0.000   & -5818.836   & -5465.481   \\
      300   &   704.623   &   696.053   &     0.759   & -5818.885   & -5463.286   \\
      400   &   830.859   &   712.748   &    44.770   & -5819.546   & -5344.548   \\
\hline
\end{tabular}
\end{center}
\end{table*}
\clearpage
\begin{table*}
\begin{center}
\caption{ The calculated thermochemical values of \textbf{Bromley's \((TiO_{2})_{9}\)}}
 \begin{tabular}{|r|r|r|r|c|c|}		\hline
 T	& \(S^{\circ}\) 	& \([-G^{\circ}-H^{\circ}(T_{r})]/T \)	& \(H^{\circ} - H^{\circ}(T_{r})\)	& \(\Delta_{f}H^{\circ}	\) 	& \(\Delta_{f}G^{\circ}\) \\
 K	& \(J mol^{-1}K^{-1}\)	& \(J mol^{-1}K^{-1}\)			& \(kJ mol^{-1}\)			& \(kJ mol^{-1}\)			& \(kJ mol^{-1}\)		\\ \hline
        0   &       0.000   &  Infinite   &   -77.878   & -6578.040   &       -6578.040   \\
      100   &   405.981   &  1079.088   &   -68.247   & -6599.595   & -6476.782   \\
      200   &   592.208   &   787.536   &   -40.433   & -6615.232   & -6347.192   \\
      298.15   &   755.288   &   748.789   &    0.000   & -6621.779   & -6213.950   \\
      300   &   758.139   &   748.796   &     0.853   & -6621.835   & -6211.418   \\
      400   &   900.586   &   767.617   &    50.522   & -6622.422   & -6074.398   \\
\hline
\end{tabular}
\end{center}
\end{table*}
\clearpage
\begin{table*}
\begin{center}
\caption{ The calculated thermochemical values of \textbf{Bromley's \((TiO_{2})_{10}\)}}
 \begin{tabular}{|r|r|r|r|c|c|}		\hline
 T	& \(S^{\circ}\) 	& \([-G^{\circ}-H^{\circ}(T_{r})]/T \)	& \(H^{\circ} - H^{\circ}(T_{r})\)	& \(\Delta_{f}H^{\circ}	\) 	& \(\Delta_{f}G^{\circ}\) \\
 K	& \(J mol^{-1}K^{-1}\)	& \(J mol^{-1}K^{-1}\)			& \(kJ mol^{-1}\)			& \(kJ mol^{-1}\)			& \(kJ mol^{-1}\)		\\ \hline
        0   &       0.000   &  Infinite   &   -88.498   & -7419.038   &       -7419.038   \\
      100   &   423.047   &  1189.738   &   -77.702   & -7442.892   & -7303.629   \\
      200   &   638.627   &   858.999   &   -45.566   & -7459.036   & -7157.337   \\
      298.15   &   822.513   &   815.436   &    0.000   & -7465.671   & -7007.549   \\
      300   &   825.711   &   815.448   &     0.956   & -7465.725   & -7004.706   \\
      400   &   984.759   &   836.484   &    56.406   & -7466.114   & -6850.842   \\
\hline
\end{tabular}
\end{center}
\end{table*}

\clearpage

\label{lastpage}


\begin{thebibliography}{}
\bibliographystyle{mn2e}

\bibitem{} Bilger C., Rimmer P.B., Helling Ch. 2013, MNRAS 435, 1888
\bibitem{} Bromley S.T., Moreira I.P.R., Neyman K.M., and Illas F 2009, Chem. Soc. Rev, 38, 2657-2670
\bibitem{} Calatayud M., Maldonado L., and Minot C 2008, J.Phys. Chem. C, 112, 16087-16095
\bibitem{} Catlow C. R. A., Bromley S. T., Hamad S., Mora-Fonz M., Sokol A. A., Woodley S. M. 2010, Phys. Chem. Chem. Phys. 12(4), 786
\bibitem{} Chase M.W, Davies C.A, Downey J.R, Frurip D.J, McDonald R.A, and  Syverud A.N 1985, NIST-JANAF thermochemical tables 1985
\bibitem{} Copperwheat C.M., Wheatley P.J., Southworth J., Bento J., Marsh T.R.  et al. 2013, MNRAS 434, 661 
\bibitem{}  Demyk K., Heijnsbergen D., Helden G., and Meijer G. 2004, A\&A 420, 547-552
\bibitem{} Dehn M., 2007, PhD thesis, Univ. Hamburg
\bibitem{} Faherty J., Rice E., Cruz K.L., Mamajek E.E., Nunez A. 2013, AJ 145, 2
\bibitem{} Frisch M.J., Trucks G.W.,  Schlegel H. B.,  Scuseria G. E.,  Robb M. A. et al. 2009, Gaussian 09 {R}evision {D}.01, Gaussian Inc. Wallingford CT 2009
\bibitem{}Gail H.P., Keller R Sedlmayr E. 1984, A\&A 133, 320
\bibitem{}Gail H.P., Sedlmayr E. 1986, A\&A 166, 225
\bibitem{}Gail H.P., Wetzel S., Pucci A., Tamanai A. 2013, A\&A 555, 119
\bibitem{} Gail H.P., Sedlmayr E. 2014, Physics and Chemistry of
  Circumstellar Dust Shells, Cambridge Astrophysics Series 52
\bibitem{}Grevesse N., Asplund M., Sauval A. J., 2007, Space Sci. Rev.,130, 105
\bibitem{} Goumans T.P.M., Bromley S.T. 2013, Phil. Trans. R. Soc. A  371 (1994), 20110580
\bibitem{}Helling Ch., Winters J.-M., Sedlmayr E. 2000, A\&A 358, 651
\bibitem{}Helling Ch., Oevermann M., L\"uttke M.J.H., Klein R., Sedlmayr E. 2001, A\&A 376, 194
\bibitem{}Helling Ch., Woitke P. 2006, A\&A 455, 325
\bibitem{}Helling Ch. 2007, Proceedings of the International Astronomical Union (2006) 2, 224
\bibitem{}Helling Ch., Woitke P., Thi W.-F. 2008a, A\&A 485, 547
\bibitem{}Helling Ch., Dehn M., Woitke P., Hauschildt P. H. 2008b, ApJ 675, L105
\bibitem{}Helling Ch., Ackerman A., Allard F., Dehn M., Hauschildt P., Homeier, D. et al. 2008, MNRAS 391, 1854 
\bibitem{}Helling Ch., Fomins A. 2013, Phil. Trans. R. Soc. A  37 (1994), 20110581-20110581
\bibitem{} Jeong K.S., Winters J.M., Sedlmayr E. Proceedings of the International Astronomical Union (1999), 99-62044, 233
\bibitem{} Jeong K.~S., Chang C., Sedlmayr E., S{\"u}lzle D. 2000, JPhB 33, 3417
\bibitem{} Jeong K.~S, Winters J.~M., Le Bertre T., Sedlmayr E. 2003, A\&A 407, 191
\bibitem{} John M., Sedlmayr E. 1997, ApSS 251, 219
\bibitem{} K\"ohler T.M., Sedlmayr E. 1997, A\&A 310, 553
\bibitem{} Lecavelier Des Etangs A., Pont F., Vidal-Madjar A., Sing D. 2008, A\&A 481, 83
\bibitem{}  {Lee} C., {Yang} W.,  {Parr} R.~G. 1988, \prb 37, 785
\bibitem{} Miller-Ricci Kempton E., Zahnle K., Fortney J.J. 2012, APJ 745, 3
\bibitem{} Patzer A.B.C, Gauger A., and Sedlmayr E. 1998, A\&A 337, 847-858
\bibitem{} Richard C., Catlow A., Bromley S.T, Hamad S., Mora-Fonz M., Sokol A.A, and Woodey S.M 2010, PCCP 12, 786-811
\bibitem{} Plane J.M.C. 2013, Phil. Trans. R. Soc. A 371 (1994), 20120335
\bibitem{} Pont F., Sing D.K., Gibson N.P., Aigrain S., Henry G., Husnoo N. 2013, MNRAS 432, 2917
\bibitem{} Sing D.K., Pont F., Aigrain S., Charbonneau D., Desert J.-M. et al. 2011, MNRAS 416, 1443 
\bibitem{} Sing D.K., Lecavelier des Etangs A., Fortney J.J., Burrows A.S., Pont F. et al. 2013, MNRAS 436, 2956
\bibitem{} Schlawin E., Zhao M., Teske J.K., Herter T. 2014, ApJ 783, 5
\bibitem{} Syzgantseva O.A, Gonzalez-Navarrete P., Calatayud M., Bromley S.T, and Minot C. 2011, J.Phys. Chem. C, 115(32), 15890-15899
\bibitem{} Wetzel S., Klevenz M., Gail H.-P., Pucci A., Trieloff M. 2012, A\&A 2013, 553, 92
\bibitem{} Witte S., Helling Ch., Hauschildt P. 2009, A\&A 506, 1367
\bibitem{} Woitke P., Helling C., 2003, A\&A, 399, 297
\bibitem{} Woitke P., Helling C., 2004, A\&A, 414, 335
\bibitem{} Gaussian 09, Frisch et al, Gaussian, Inc., Wallingford CT, 2009.
\bibitem{}  Lee C., Yang W., Parr R.G 1988, Phys. Rev. B 37  785-789
\end{thebibliography}
\end{document}